\documentclass[useAMS,usenatbib,usegraphicx,referee]{mn2e}
\usepackage{graphicx}
\usepackage{color}
\newcounter{subfigure}
\begin{document}
%
%
\title[The SAURON project - XIII]{The SAURON project - XIII. SAURON-GALEX study
of early-type galaxies: the ultraviolet colour-magnitude relations
and Fundamental Planes}
\author[H.\ Jeong et al.]{Hyunjin Jeong,$^{1}$ Sukyoung K. Yi,$^{1}$\thanks{E-mail: yi@yonsei.ac.kr}
 Martin Bureau,$^{2}$ Roger L. Davies,$^{2}$
\newauthor Jes\'{u}s Falc\'{o}n-Barroso,$^{3,4}$ Glenn van de Ven,$^{5}$\thanks{Hubble Fellow}
 Reynier F. Peletier,$^{6}$ Roland Bacon,$^{7}$
\newauthor Michele Cappellari,$^{2}$ Tim de Zeeuw,$^{8,9}$ Eric Emsellem,$^{7}$
Davor Krajnovi\'{c},$^{2}$
\newauthor Harald Kuntschner,$^{10}$ Richard M. McDermid,$^{11}$ Marc Sarzi,$^{12}$
\newauthor and Remco C. E. van den Bosch$^{13}$ \\
$^1$Department of Astronomy, Yonsei University, Seoul 120-749, Korea\\
$^2$Sub-Department of Astrophysics, University of Oxford, Denys
Wilkinson Building, Keble Road, Oxford OX1 3RH, United Kingdom\\
$^3$European Space Agency / ESTEC, Keplerlaan 1, 2200 AG Noordwijk, The Netherlands\\
$^4$Instituto de Astrof\'isica de Canarias, E-38200 La Laguna, Tenerife, Spain\\
$^5$Institute for Advanced Study, Einstein Drive, Princeton, NJ 08540, U.S.A.\\
$^6$Kapteyn Astronomical Institute, University of Groningen, P.O. Box
800, 9700 AV Groningen, The Netherlands\\
$^7$Universit\'{e} de Lyon 1, CRAL, Observatoire de Lyon, 9
av. Charles Andr\'{e}, F-69230 Saint-Genis Laval; CNRS, UMR 5574;\\
ENS de Lyon, France\\
$^8$European Souther Observatory, Karl-Schwarzchild-Str. 2, 85748, Garching, Germany\\
$^9$Leiden Observatory, Leiden University, Niels Bohrweg 2, 2333 CA Leiden, The Netherlands\\
$^{10}$Space Telescope European Coordinating Facility, European
Southern Observatory, Karl-Schwarzchild-Str. 2, 85748, Garching,
Germany\\
$^{11}$Gemini Observatory, 670 North A'Ohoku Place, Hilo, Hawaii 96720, U.S.A.\\
$^{12}$Centre for Astrophysics Research, University of Hertfordshire,
Hatfield, United Kingdom\\
$^{13}$Department of Astronomy, University of Texas, Austin, TX 78712,
U.S.A.}
\maketitle
%
%
\begin{abstract}
  We present {\it Galaxy Evolution Explorer} ({\it GALEX}) far (FUV)
  and near (NUV) ultraviolet imaging of $34$ nearby early-type
  galaxies from the {\tt SAURON} representative sample of $48$ E/S0
  galaxies, all of which have ground-based optical imaging from the
  MDM Observatory. The surface brightness profiles of nine galaxies
  ($\approx26$~per cent) show regions with blue UV$-$optical colours
  suggesting recent star formation. Five of these ($\approx15$~per
  cent) show blue integrated UV$-$optical colours that set them aside
  in the NUV integrated colour-magnitude relation. These are objects
  with either exceptionally intense and localised NUV fluxes or blue
  UV$-$optical colours throughout. They also have other properties
  confirming they have had recent star formation, in particular
  H$\beta$ absorption higher than expected for a quiescent population
  and a higher CO detection rate. This suggests that residual star
  formation is more common in early-type galaxies than we are used to
  believe. NUV-blue galaxies are generally drawn from the lower
  stellar velocity dispersion ($\sigma_{\rm e}<200$~km~s$^{-1}$) and
  thus lower dynamical mass part of the sample. We have also
  constructed the first UV Fundamental Planes and show that NUV blue
  galaxies bias the slopes and increase the scatters. If they are
  eliminated the fits get closer to expectations from the virial
  theorem. Although our analysis is based on a limited sample, it
  seems that a dominant fraction of the tilt and scatter of the UV
  Fundamental Planes is due to the presence of young stars in
  preferentially low-mass early-type galaxies. Interestingly, the
  UV$-$optical radial colour profiles reveal a variety of behaviours,
  with many galaxies showing signs of recent star formation, a central
  UV-upturn phenomenon, smooth but large-scale age and metallicity
  gradients, and in many cases a combination of these. In addition,
  FUV$-$NUV and FUV$-V$ colours even bluer than those normally
  associated with UV-upturn galaxies are observed at the centre of
  some quiescent galaxies. Four out of the five UV-upturn galaxies
  are slow rotators. These objects should thus pose interesting
  challenges to stellar evolutionary models of the UV-upturn.

\end{abstract}
%
%
\section{INTRODUCTION}
\label{sec:intro}
The dominant formation mechanism of early-type galaxies remains one
of the long-standing debates of modern astrophysics. The classical
monolithic model \citep*[e.g.][]{els62,la74} suggests that
early-type galaxies form in highly efficient starbursts at high
redshifts and evolve without much residual star formation from that
point onward. On the other hand, the popular lambda cold dark matter
($\Lambda$CDM) paradigm strongly suggests a hierarchical merger
scenario for massive early-type galaxies \citep[e.g.][]{tt72}. In
this model, early-type galaxies form as the result of successive
mergers and are thought to have continued or episodic star formation
events.

Numerous tests discriminate between these scenarios. For example, the
optical colour-magnitude relations (CMRs) of early-type galaxies
consistently reveal a small scatter around the mean relation
\citep*[e.g.][]{betal92,eetal97,setal98,vetal00}, apparently
supporting the monolithic scenario. Deep imaging surveys have however
shown that many early-types possess shells, tidal features
\citep*[e.g.][]{ss92} and signatures of ongoing or recent star
formation \citep*[RSF; e.g. NGC~2865,][]{hcb99,retal05}.
Integral-field spectroscopy now also allows to obtain spatially
resolved maps of various galaxy properties. For example, we have
previously highlighted the importance of central stellar discs and
kinematically-decoupled cores in the {\tt SAURON} sample
(e.g., \citealt*{zetal02}, hereafter \citeauthor{zetal02};
\citealt*{ketal08}, hereafter \citeauthor{ketal08}).

More recently, the {\it Galaxy Evolution Explorer} ({\it GALEX})
satellite opened up the ultraviolet (UV) window, allowing to probe
the recent star formation history of galaxies with much greater
accuracy than was previously possible with only optical information.
The rest-frame UV is one order of magnitude more sensitive than the
optical to the presence of hot stars, easily revealing populations
younger than $1$~Gyr.
\citet{yetal05} used {\it GALEX} Medium Imaging Survey data to
construct the first near ultraviolet (NUV) CMR of early-type galaxies
classified by the Sloan Digital Sky Survey (SDSS). They found a
remarkably large scatter towards blue colours and interpreted it as
evidence of recent star formation. \citet{ketal07} and
\citet{scetal07} subsequently found that the fraction of early-type
galaxies that experienced recent star formation in the last billion
years can be greater than $30$~per cent. \citet*{ybl09} further
concluded that the star formation rates of nearby early-type galaxies
derived from mid-infrared imaging are in good agreement with the UV
results. There is thus clear evidence for residual star formation in
the local early-type galaxy population and for its influence on global
scaling relations, at least colour-magnitude relations.

The Fundamental Plane (FP) is another key scaling relation of
early-type galaxies, a two-dimensional plane in the three-dimensional
manifold of their global structural parameters (effective radius,
surface brightness and stellar velocity dispersion; e.g.\
\citealt{detal87,dd87}), and may also be affected by RSF. Under the
assumption of structural homology (i.e.\ that all early-type galaxies
have the same mass distribution and kinematics), the virial theorem
implies that the FP parameters should scale in a specific manner, but
observations reveal a {\em tilt} of the FP away from the virial
prediction. Numerous studies have investigated this, and the observed
tilt has been attributed to a variation of the mass-to-light ratio
across the sequence of early-type galaxies and/or to the breaking of
the homology assumption (see, e.g.,
\citealt{detal87,dd87,glb93,jetal96,sco97,petal98a,ketal00,beetal03,tretal06};
\citealt*{cetal06}, hereafter \citeauthor{cetal06};
\citealt{boetal08}). Nevertheless, much debate still exists on the
origin of the tilt and the scatter about the mean relation (e.g.\
\citealt{doetal06} and references therein).

Spatially-resolved galaxy data provide valuable information on the
details of the stellar population distributions, especially when
combined with UV data. Hence, we aim here to obtain a UV database
matching the {\tt SAURON} integral-field data, and present
spatially-resolved UV and optical imaging of most early-type galaxies
in the {\tt SAURON} survey.  Numerous investigations can be performed
with such a database, but we will focus here on revisiting the effects
of star formation on the scaling relations of early-types,
particularly colour-magnitude relations and the Fundamental
Plane. Longstanding enigmas such as the UV-upturn phenomenon (see
\citealt{oco99} and references therein) are closely related.

In this paper, we thus present and discuss our UV imaging of {\tt
SAURON} early-type galaxies obtained with {\it GALEX}, along with
ground-based optical imaging from the MDM Observatory. In
\S~\ref{sec:obs}, we present a brief summary of the {\tt SAURON}
survey and our optical observations, and then describe at length the
{\it GALEX} observations and surface photometry. In
\S~\ref{sec:discussion}, we discuss the UV colour-magnitude
relations, the first UV Fundamental Planes, and make an attempt at
interpreting radial UV$-$optical colour profiles. We summarize our
findings in \S~\ref{sec:conclusions}.
%
%
\section{OBSERVATIONS AND DATA REDUCTION}
\label{sec:obs}
%
%
\subsection{SAURON integral-field spectroscopy}
\label{sec:sauron}
The {\tt SAURON} observations are aimed at determining the
two-dimensional stellar kinematics (\citealt*{eetal04}, hereafter
\citeauthor{eetal04}), stellar linestrengths (\citealt*{ketal06},
hereafter \citeauthor{ketal06}) and ionised gas kinematics
(\citealt*{setal06}, hereafter \citeauthor{setal06}) of $48$ nearby
early-type galaxies and $24$ spiral bulges in the field and clusters
(see \citeauthor{zetal02}). The observations and data reduction are
described in the relevant papers. Detailed dynamical modeling is
also available (see, e.g., \citeauthor{cetal06}; \citealt*{eetal07},
hereafter \citeauthor{eetal07}; \citealt*{cetal07}, hereafter
\citeauthor{cetal07}).

\renewcommand{\thefigure}{\arabic{figure}\alph{subfigure}}
%
%
\begin{figure*}
\begin{center}
\vspace*{10mm}
\includegraphics[width=7.7cm]{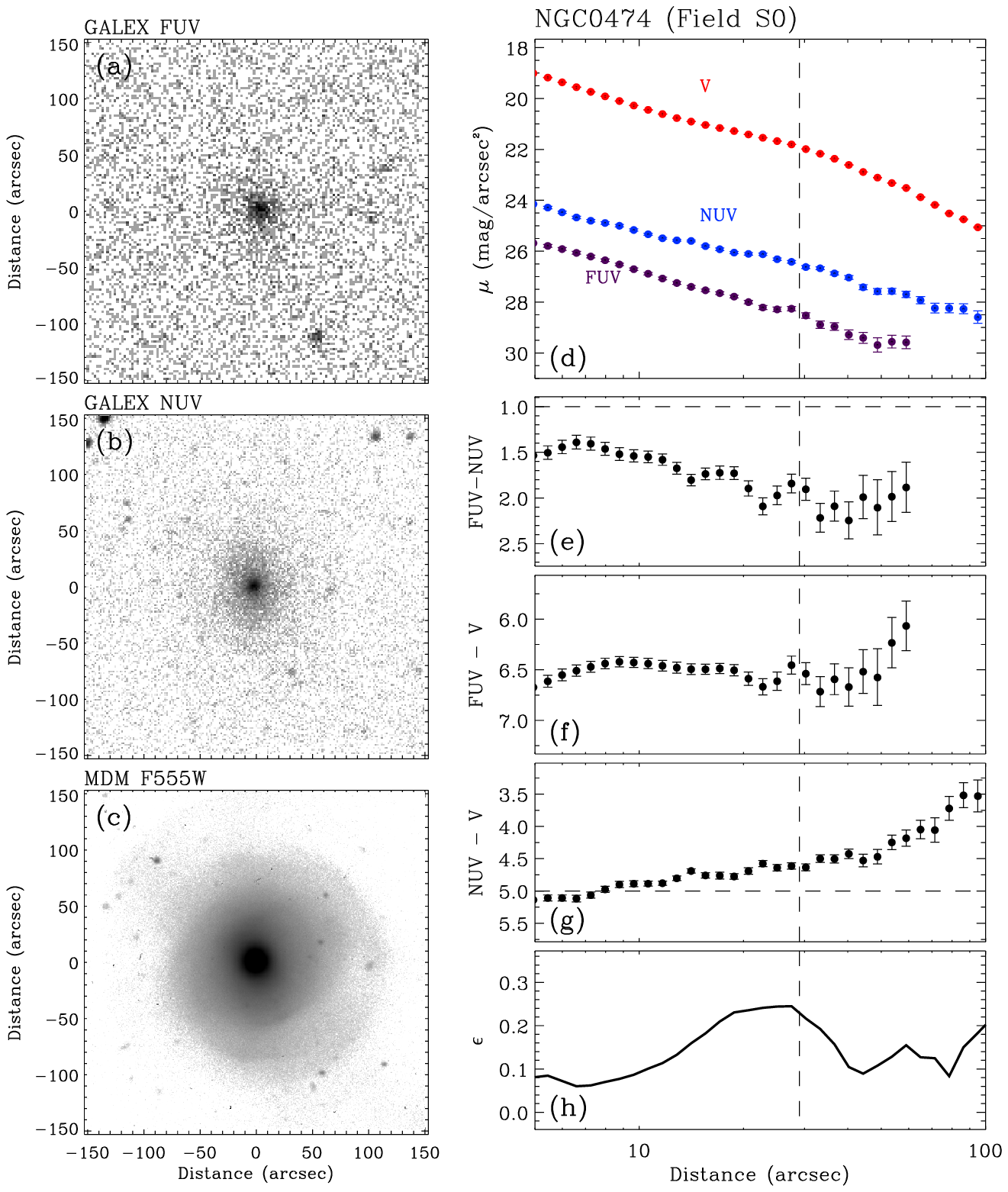}
\vspace*{10mm}
\includegraphics[width=7.7cm]{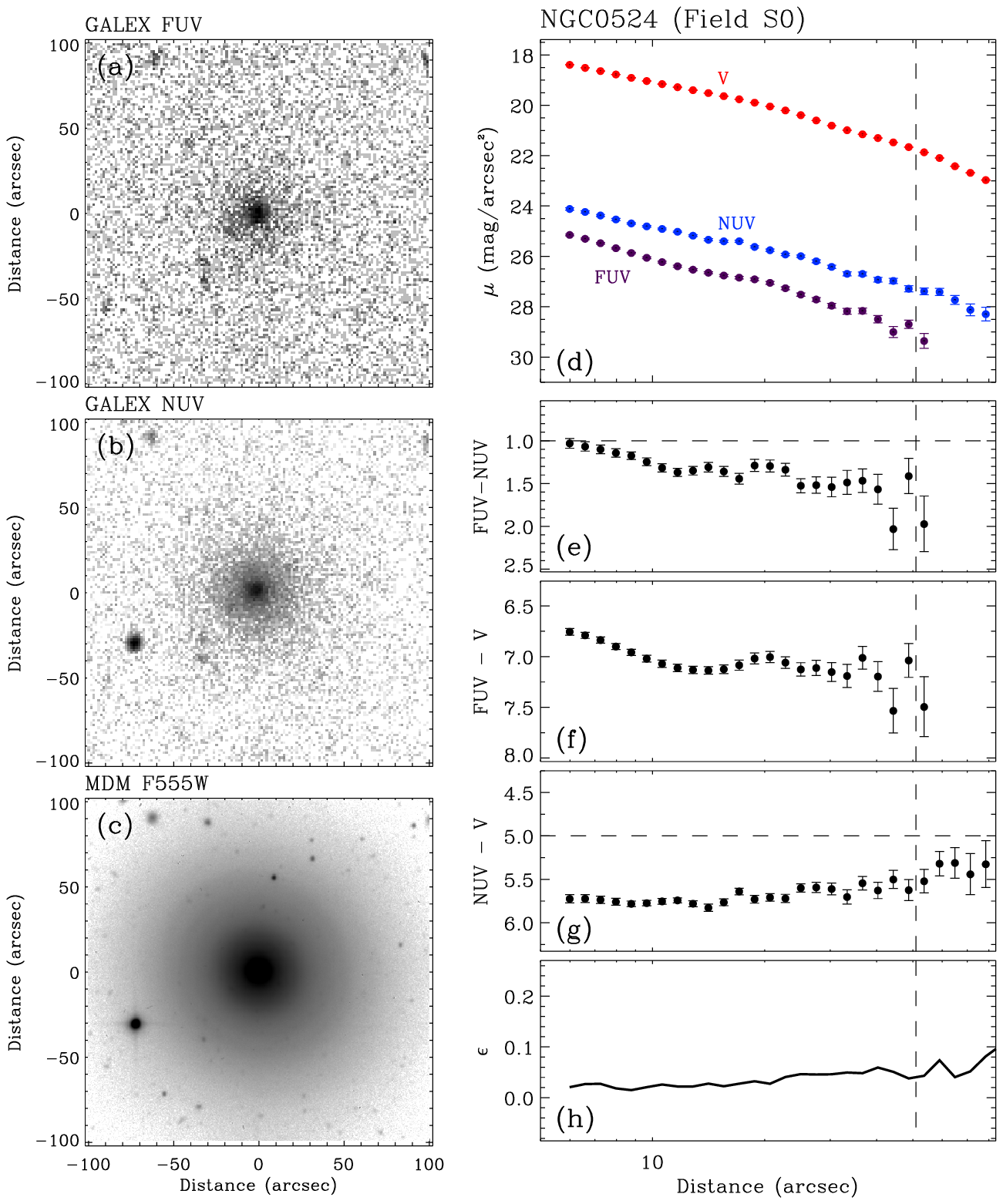}\\
\vspace*{5mm}
\includegraphics[width=7.7cm]{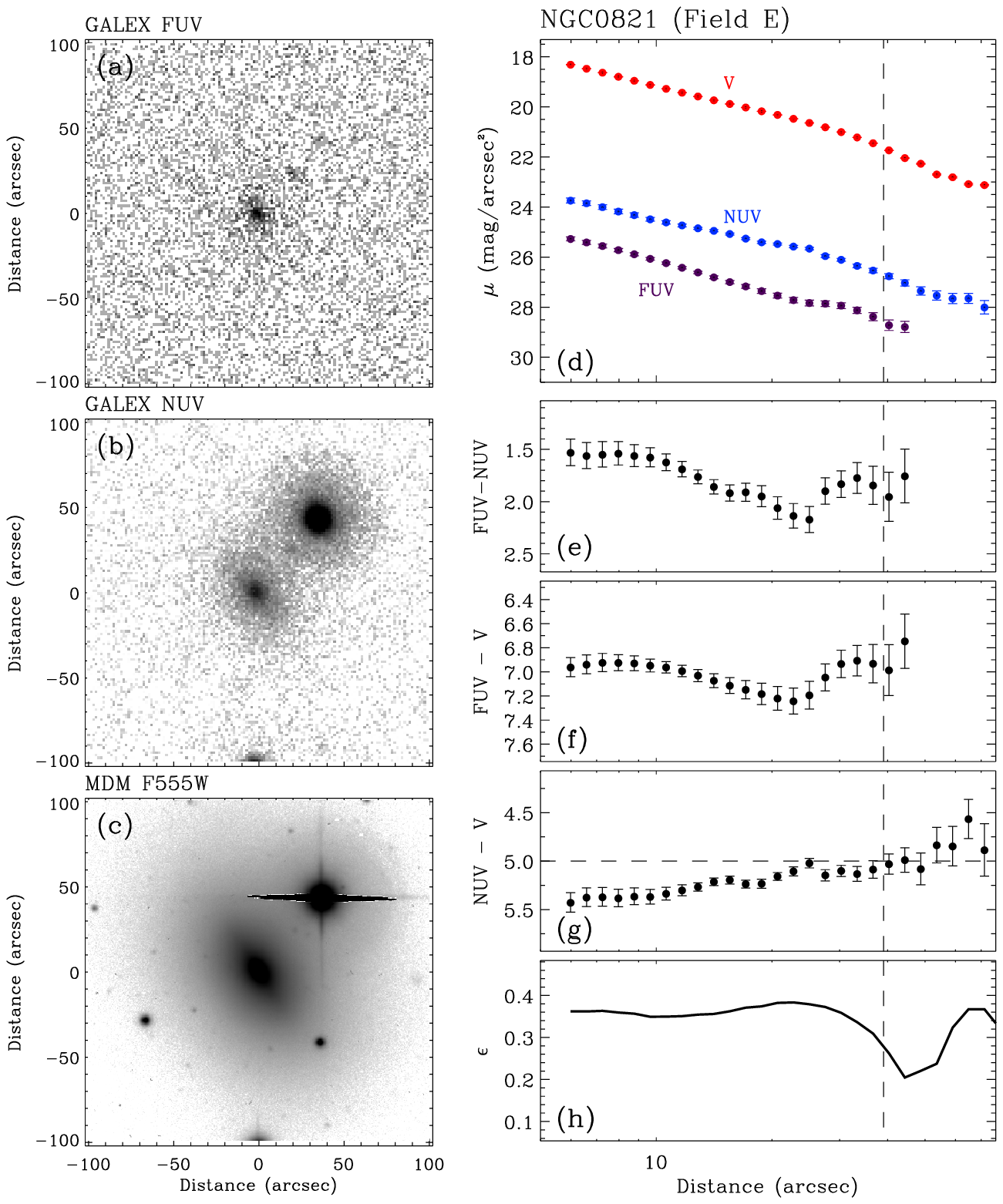}
\vspace*{5mm}
\includegraphics[width=7.7cm]{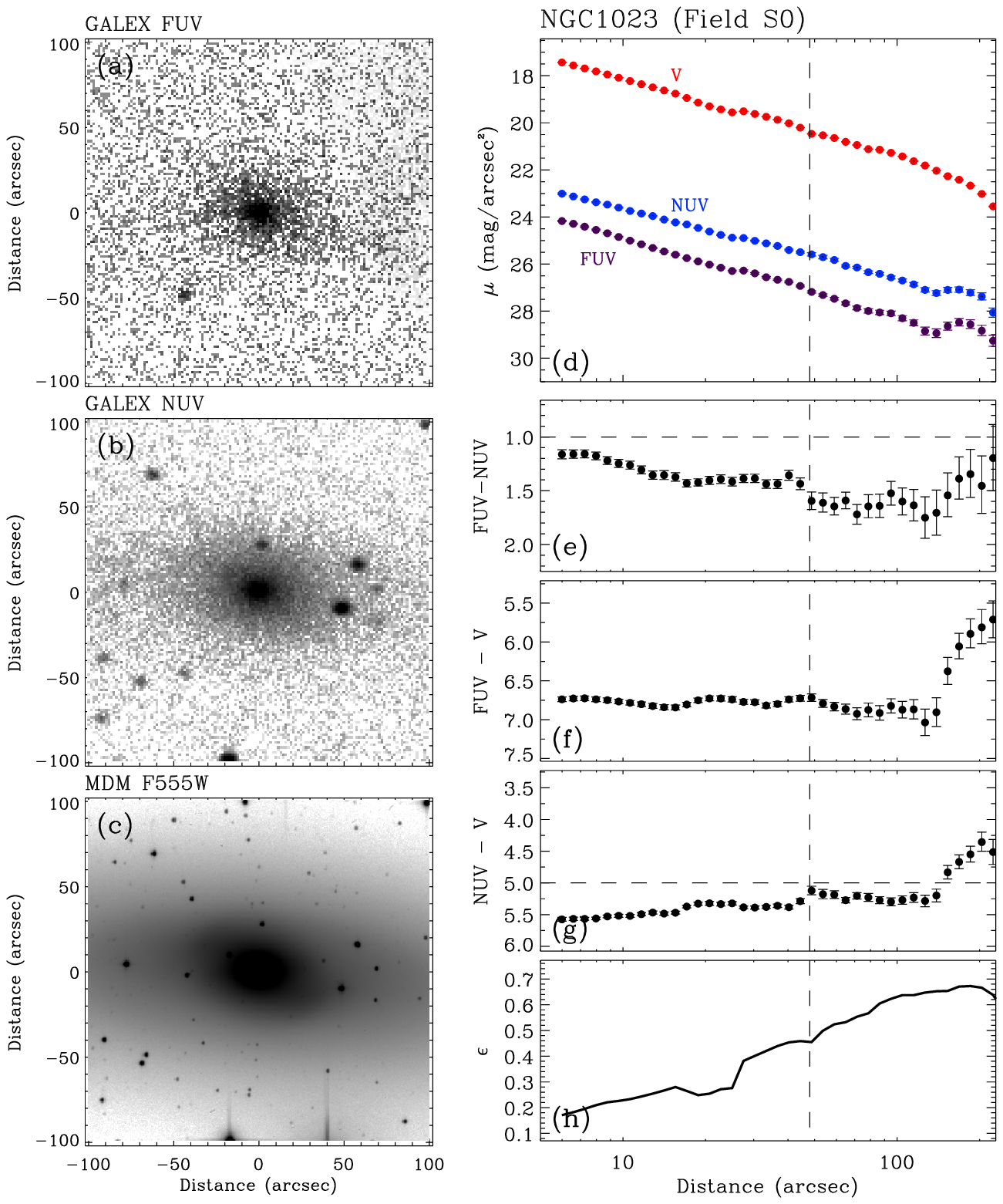}
\end{center}
\caption[]{{\it GALEX} UV and MDM optical images and surface
photometry of $34$ early-type galaxies in the {\tt SAURON} sample.
{\em (a)--(c)} FUV, NUV and F555W images. {\em (d)} $V$, NUV and FUV
radial surface brightness profiles. {\em (e)--(g)} FUV$-$NUV,
FUV$-V$ and NUV$-V$ radial colour profiles. {\em (h)} Ellipticity
radial profile derived from the optical image. The two horizontal
lines show FUV$-$NUV$\,=\,1.0$ and NUV$-V\,=\,5.0$, respectively.
The vertical lines show the $I$-band effective radius from the {\tt
SAURON} survey (\citeauthor[see][]{ketal06}).} \label{fig:results}
\end{figure*}
\addtocounter{figure}{-1}
\begin{figure*}
\begin{center}
\vspace*{10mm}
\includegraphics[width=7.7cm]{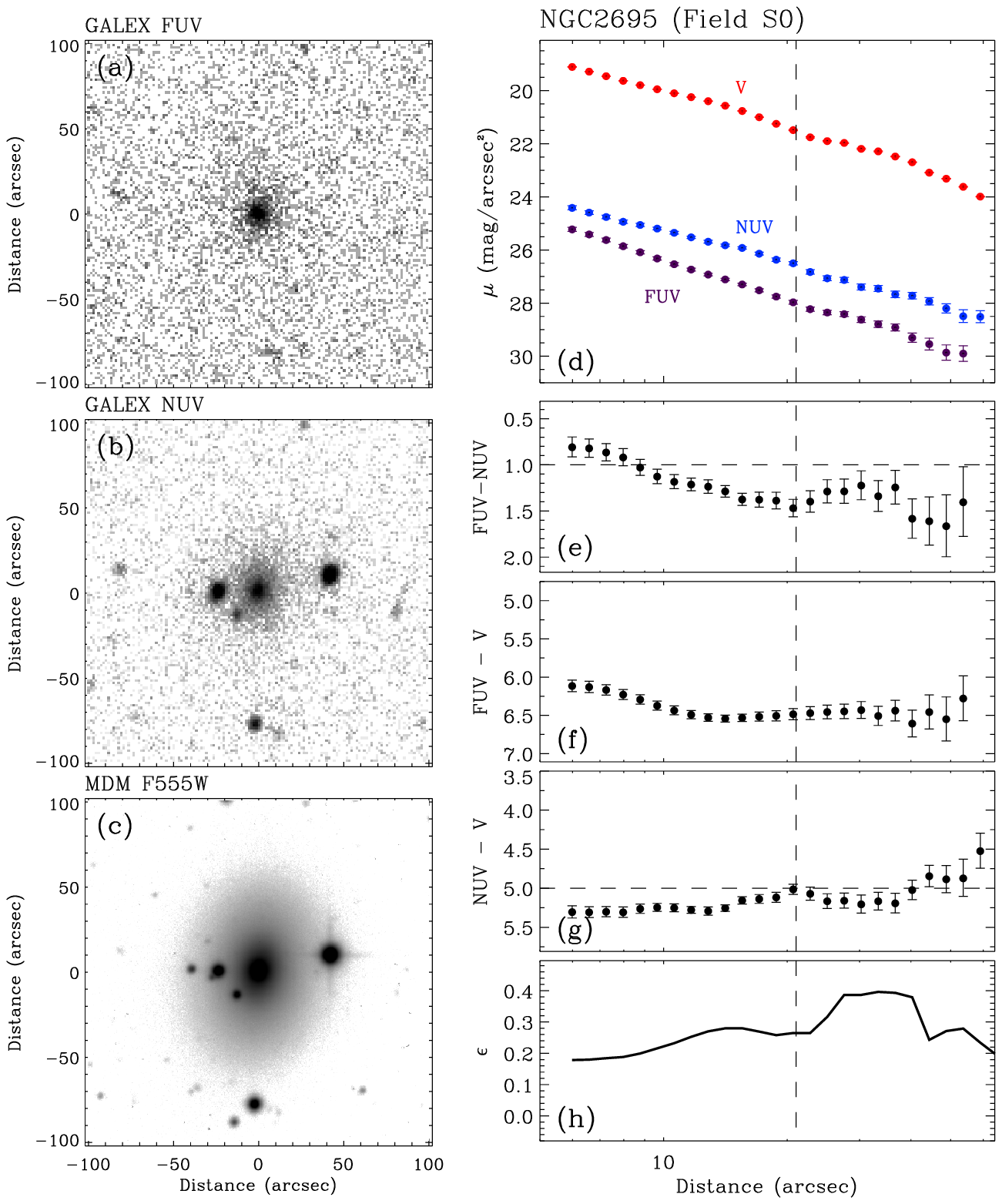}
\vspace*{10mm}
\includegraphics[width=7.7cm]{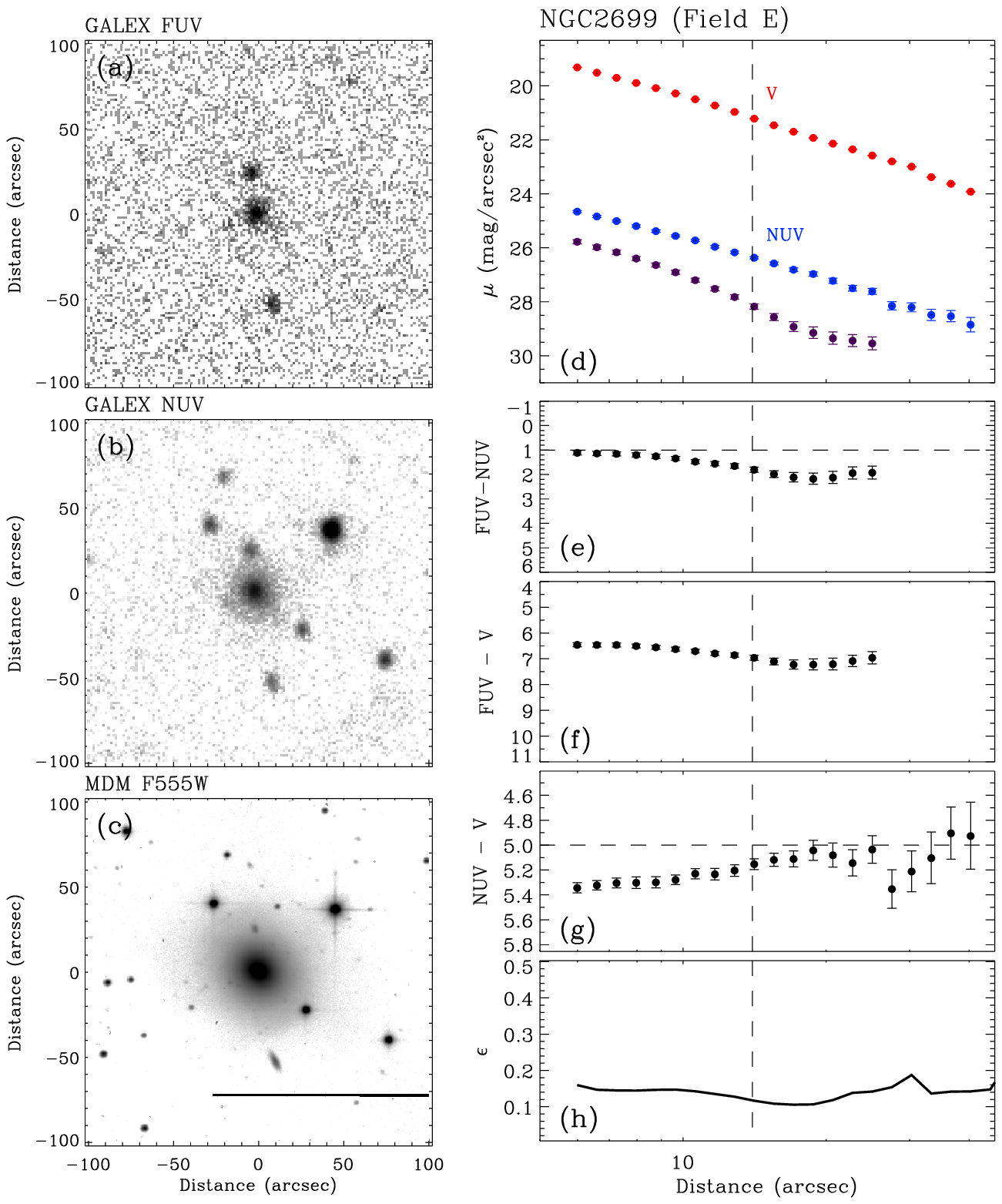}\\
\vspace*{10mm}
\includegraphics[width=7.7cm]{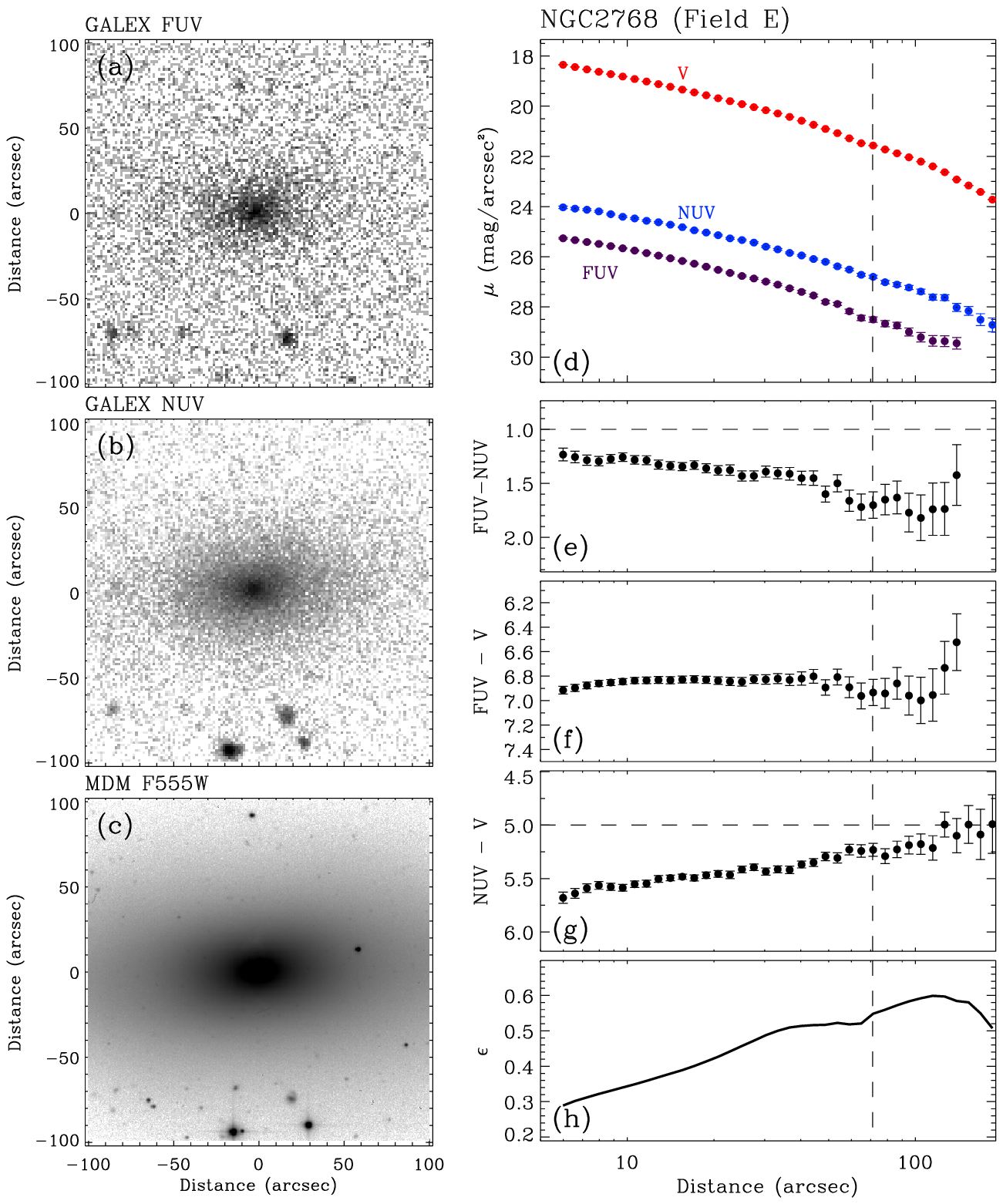}
\vspace*{10mm}
\includegraphics[width=7.7cm]{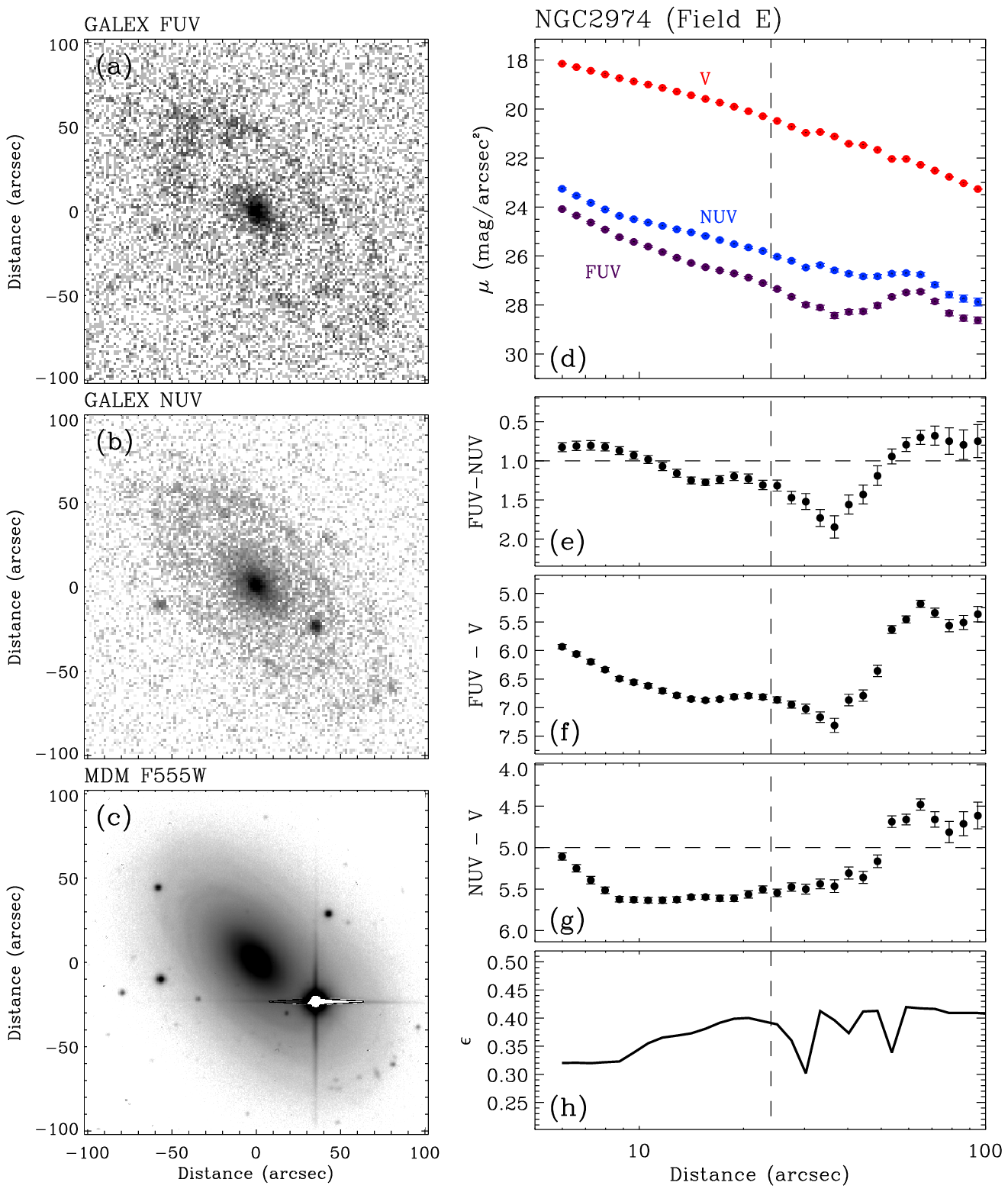}
\end{center}
\caption[]{continued.}
\end{figure*}
\addtocounter{figure}{-1}
\begin{figure*}
\begin{center}
\vspace*{10mm}
\includegraphics[width=7.7cm]{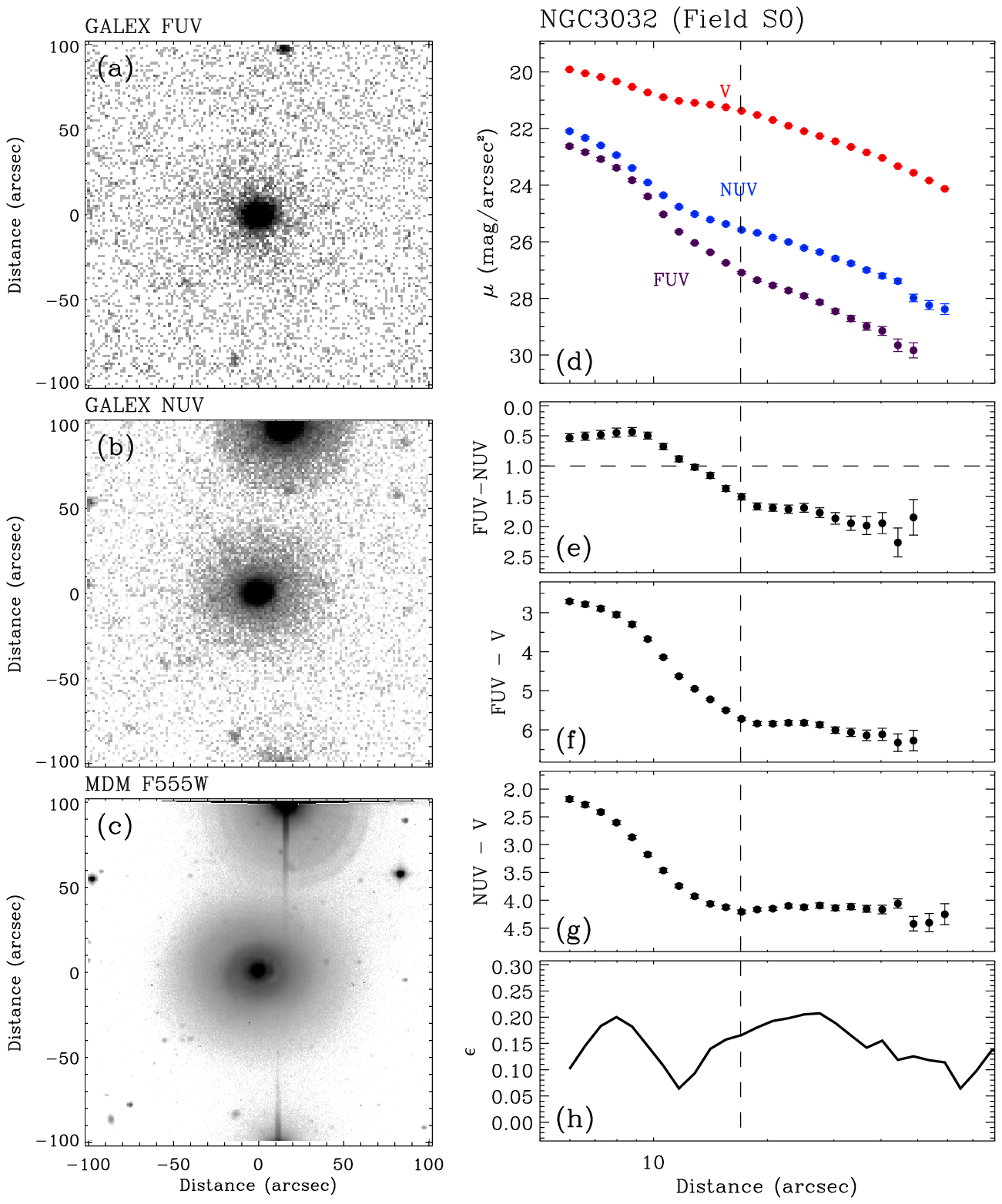}
\vspace*{10mm}
\includegraphics[width=7.7cm]{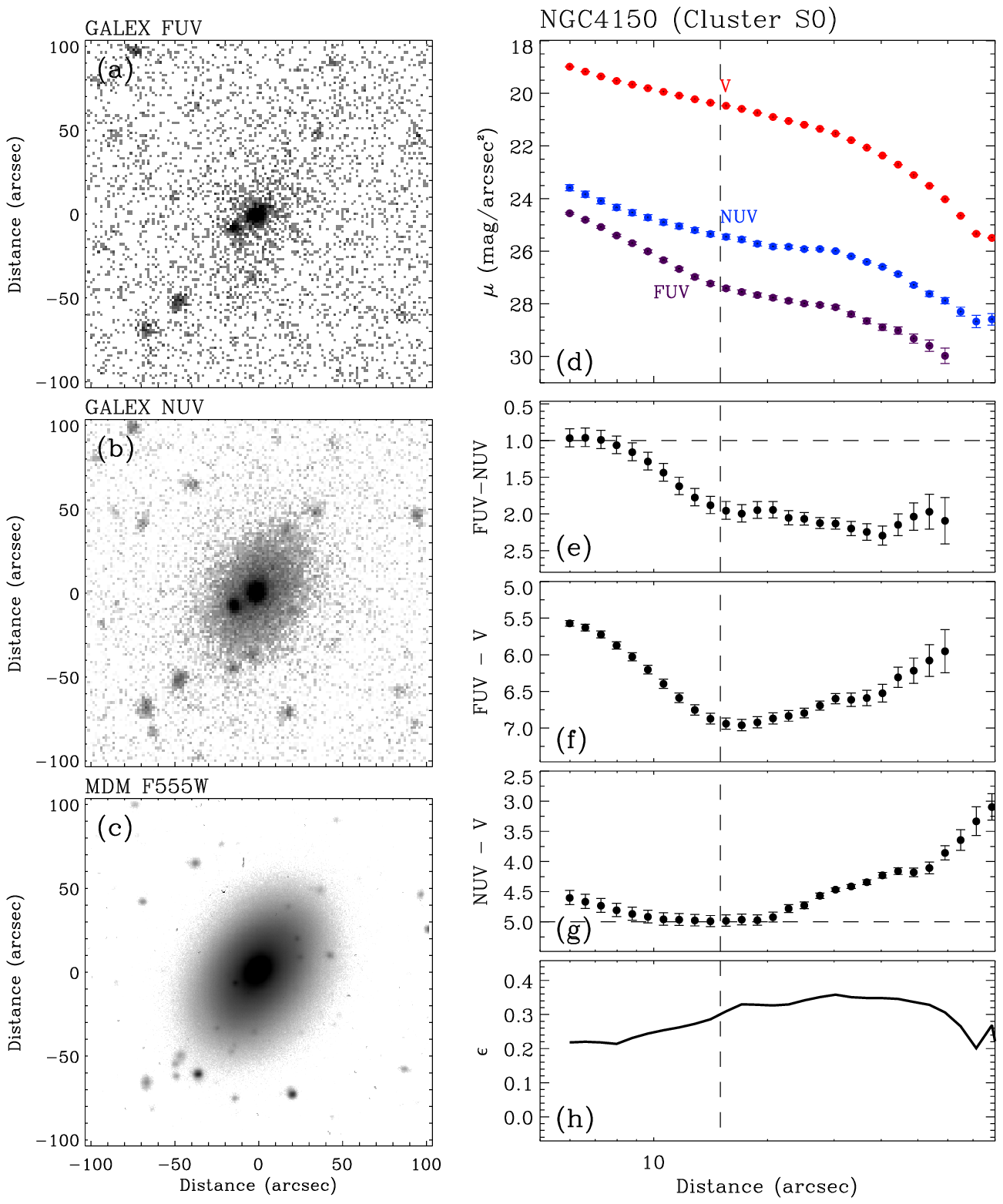}\\
\vspace*{10mm}
\includegraphics[width=7.7cm]{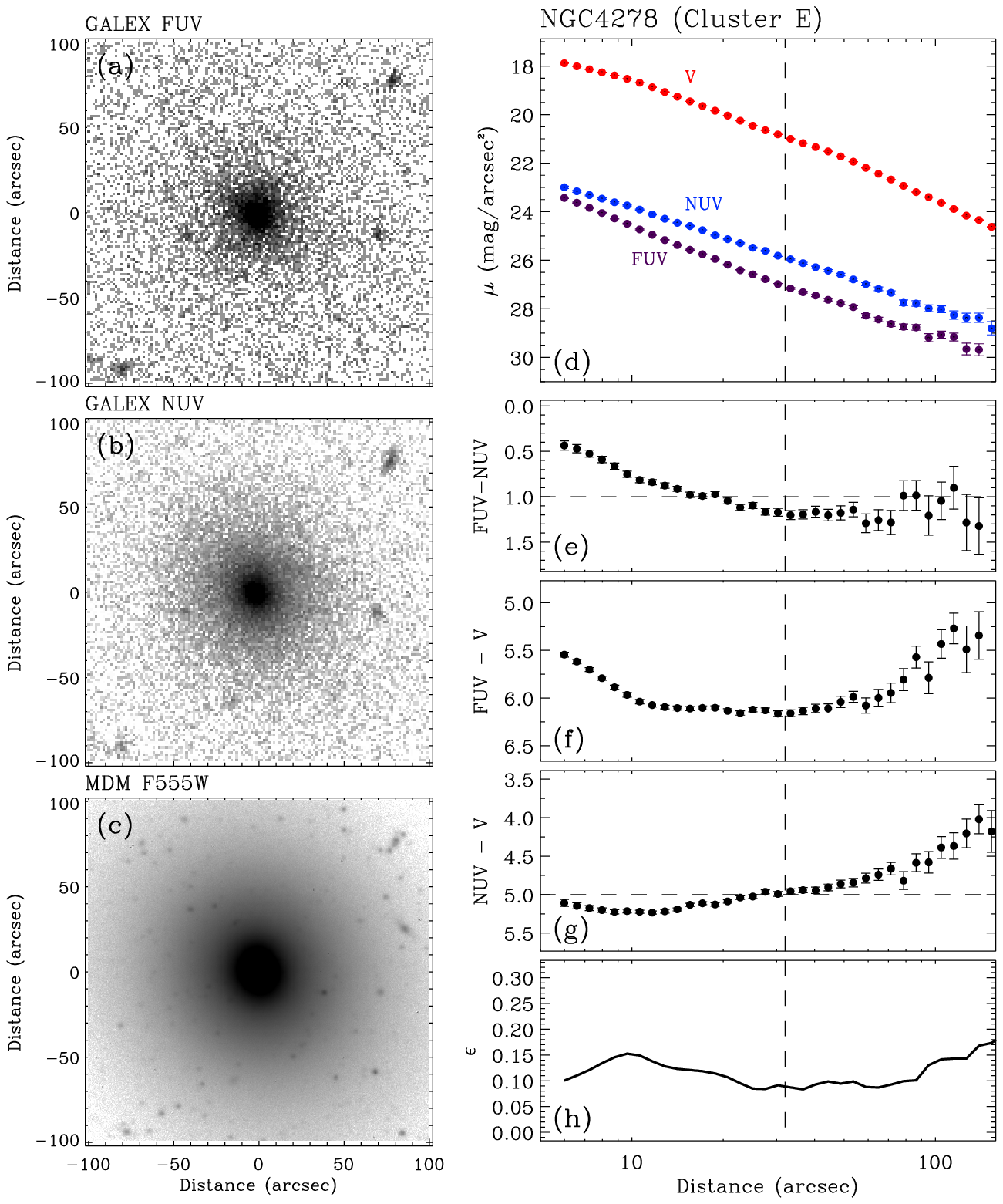}
\vspace*{10mm}
\includegraphics[width=7.7cm]{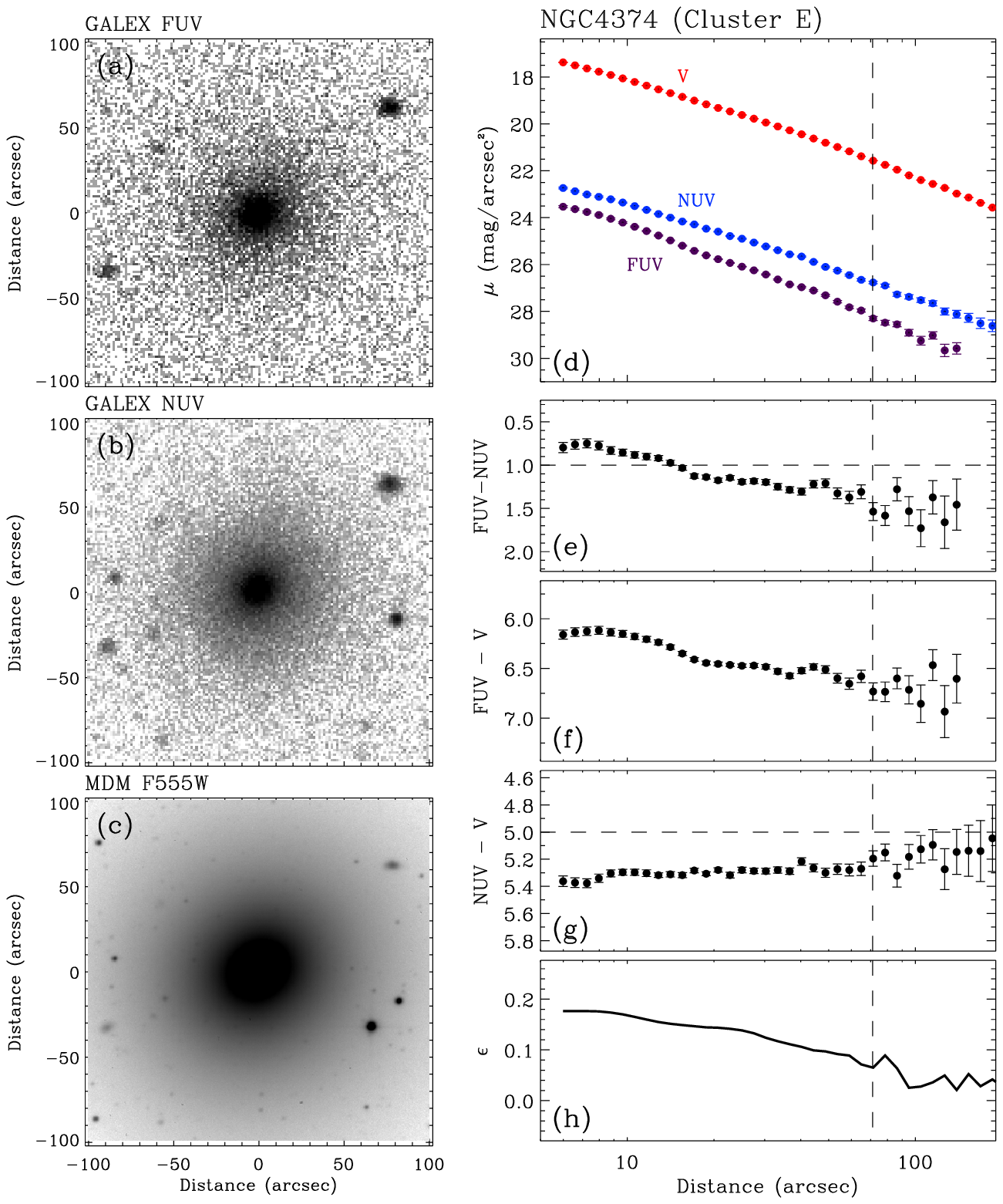}
\end{center}
\caption[]{continued.}
\end{figure*}
\addtocounter{figure}{-1}
\begin{figure*}
\begin{center}
\vspace*{10mm}
\includegraphics[width=7.7cm]{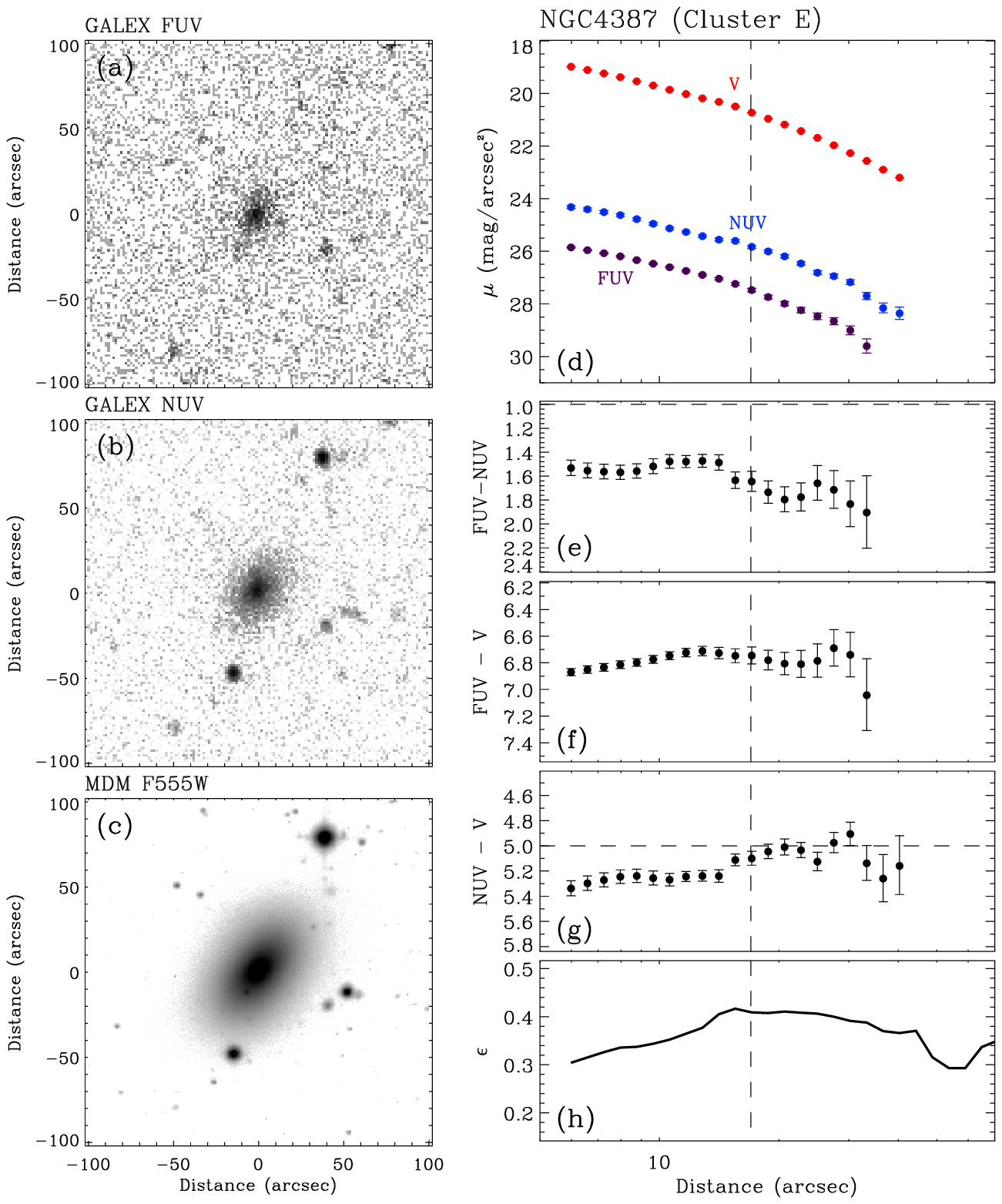}
\vspace*{10mm}
\includegraphics[width=7.7cm]{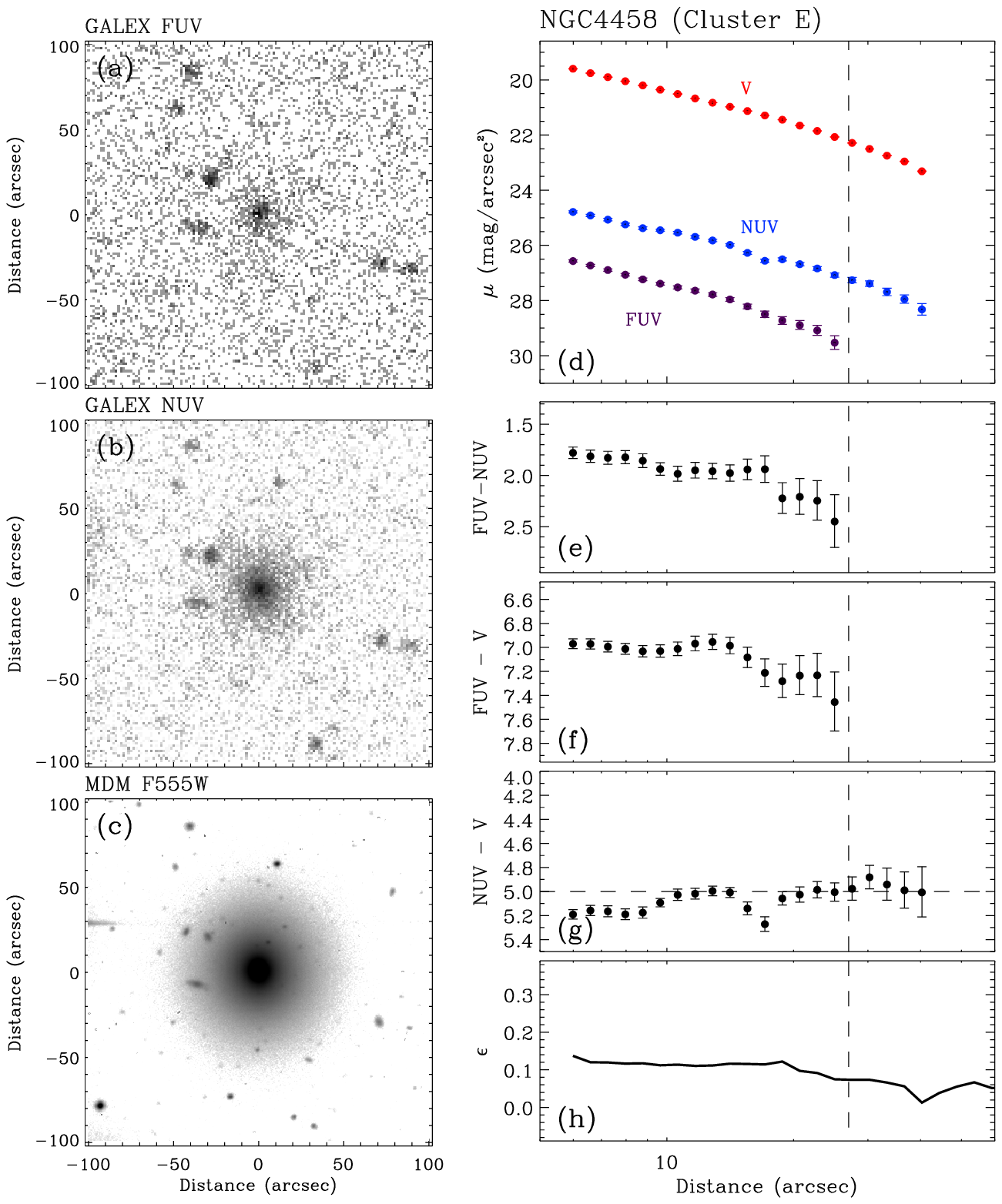}\\
\vspace*{10mm}
\includegraphics[width=7.7cm]{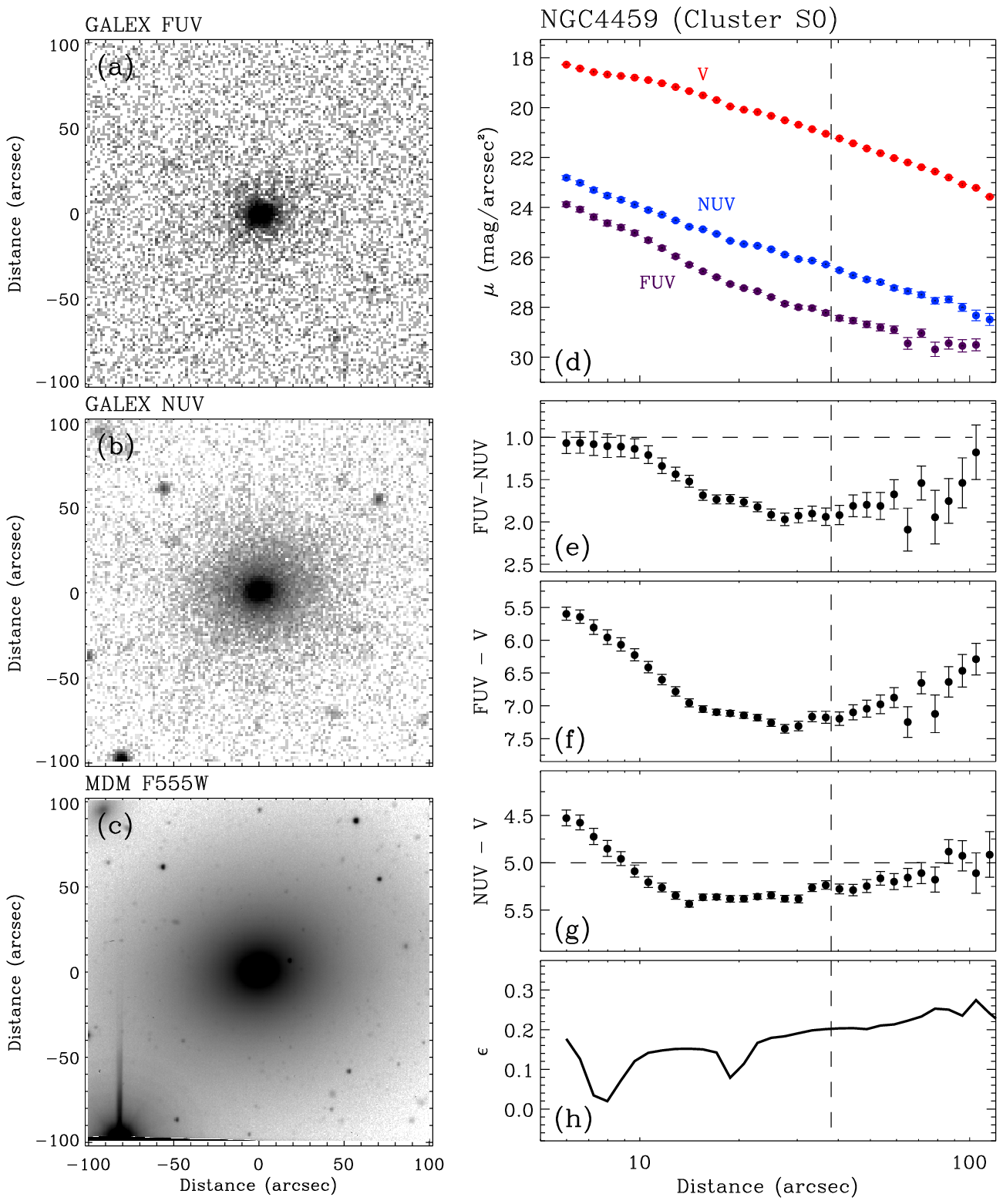}
\vspace*{10mm}
\includegraphics[width=7.7cm]{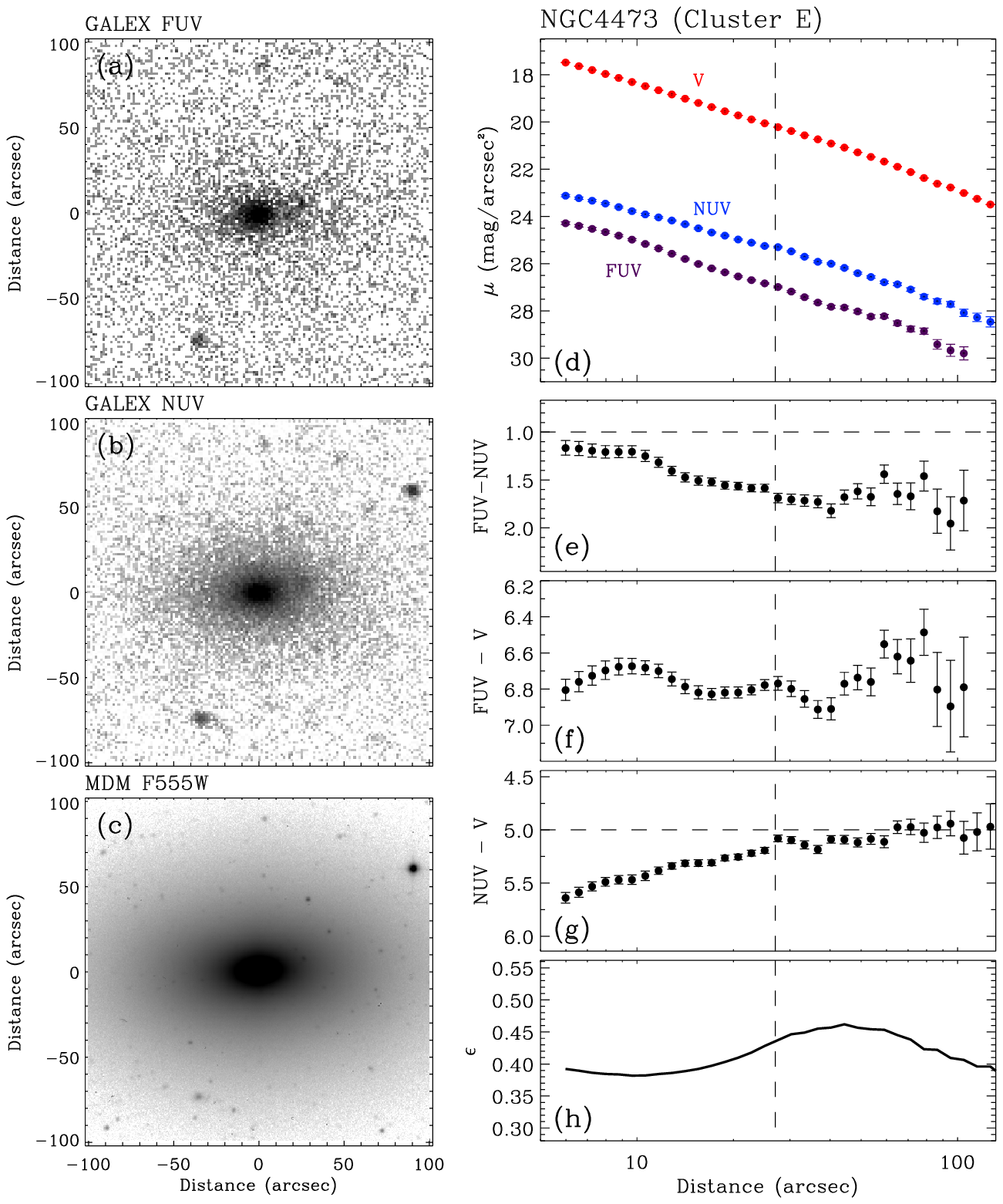}
\end{center}
\caption[]{continued.}
\end{figure*}
\addtocounter{figure}{-1}
\begin{figure*}
\begin{center}
\vspace*{10mm}
\includegraphics[width=7.7cm]{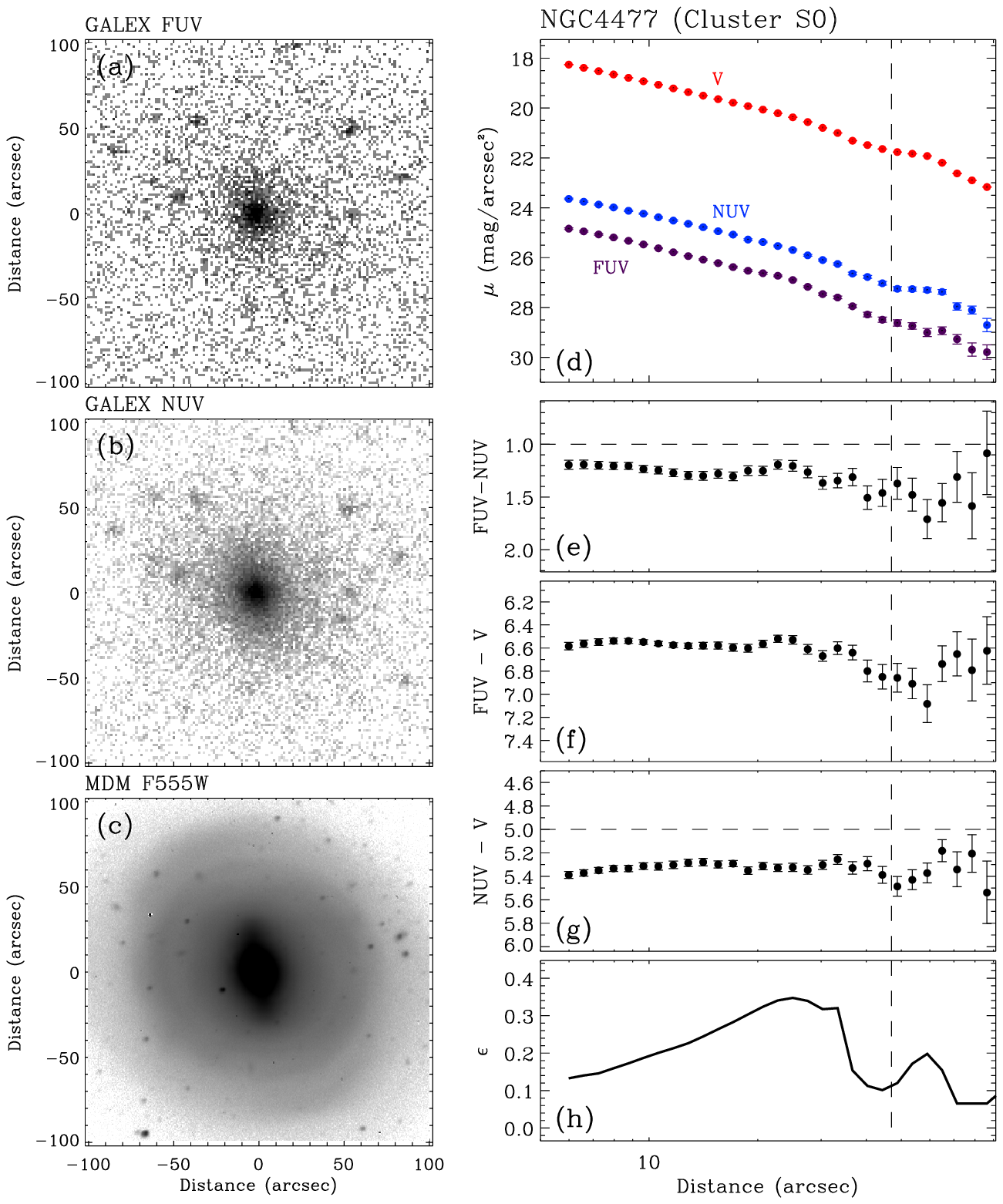}
\vspace*{10mm}
\includegraphics[width=7.7cm]{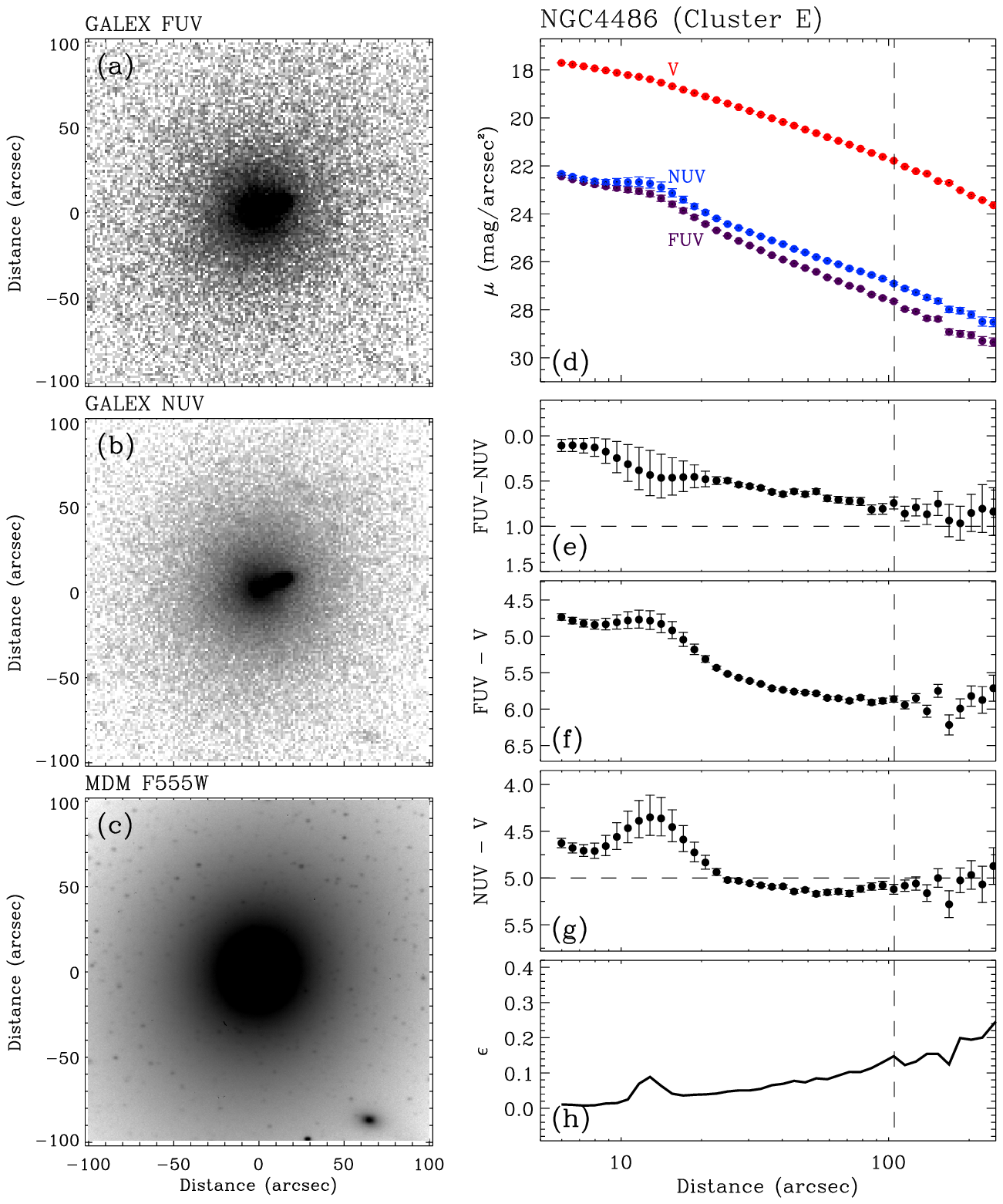}\\
\vspace*{10mm}
\includegraphics[width=7.7cm]{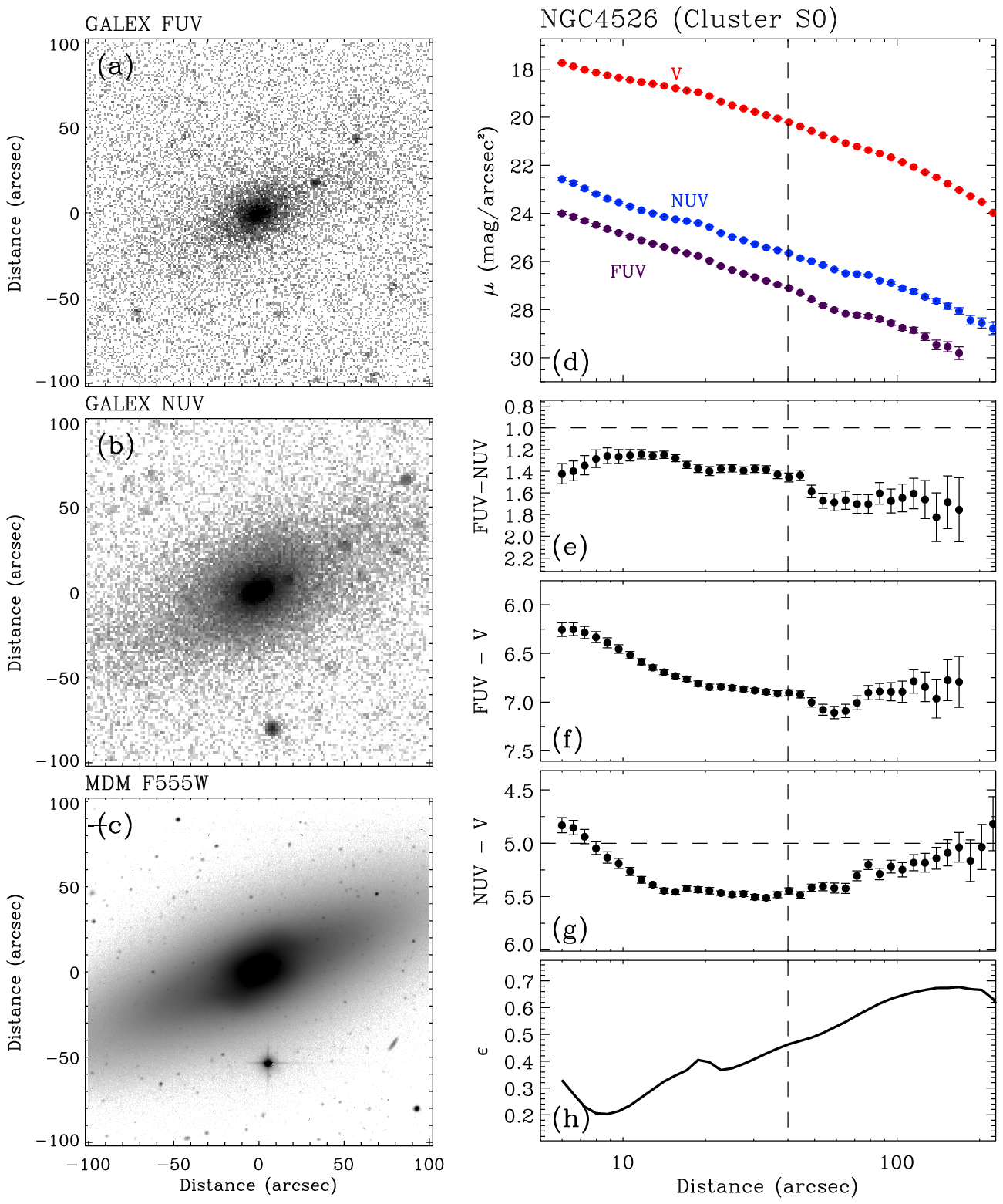}
\vspace*{10mm}
\includegraphics[width=7.7cm]{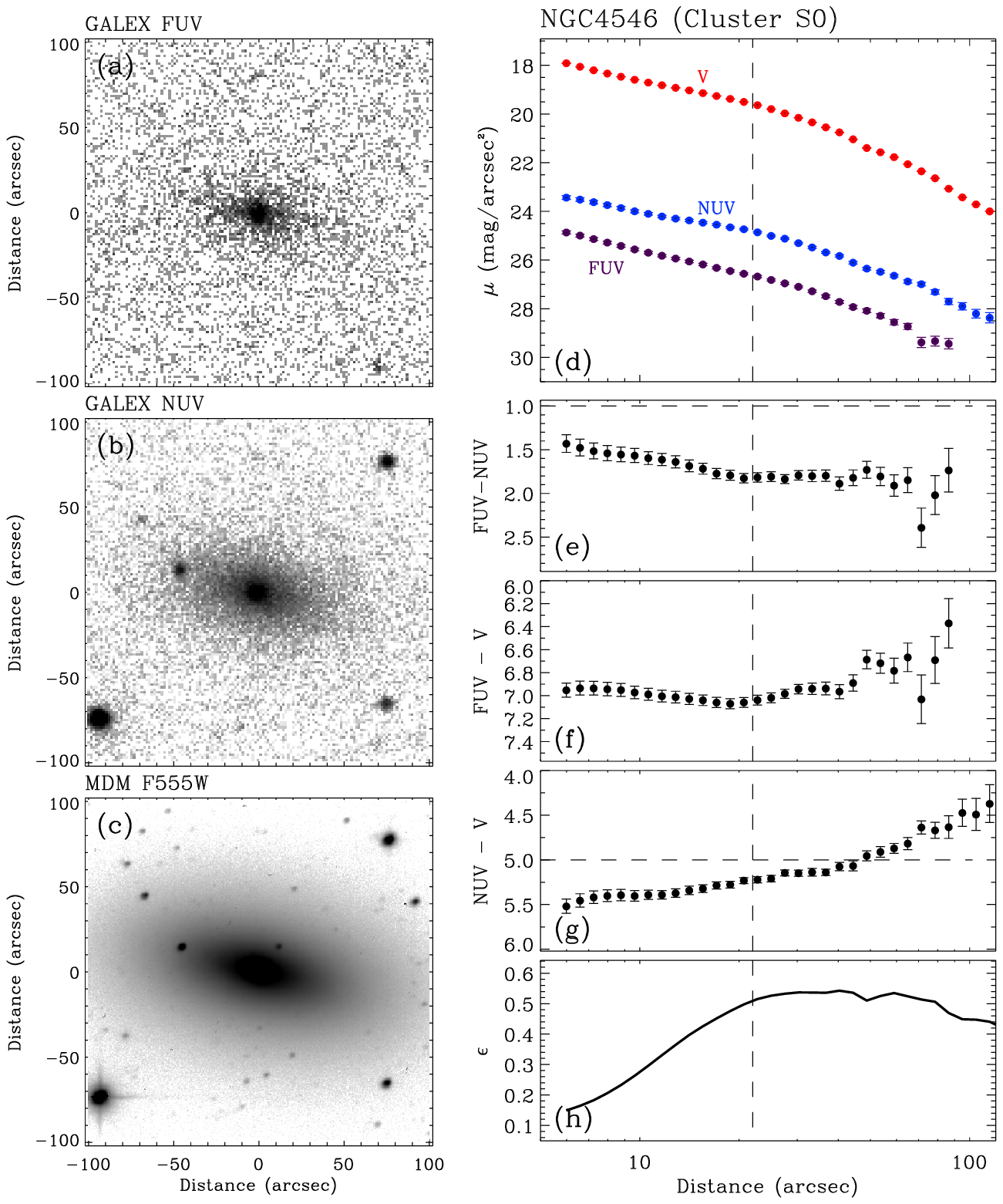}
\end{center}
\caption[]{continued.}
\end{figure*}
\addtocounter{figure}{-1}
\begin{figure*}
\begin{center}
\vspace*{10mm}
\includegraphics[width=7.7cm]{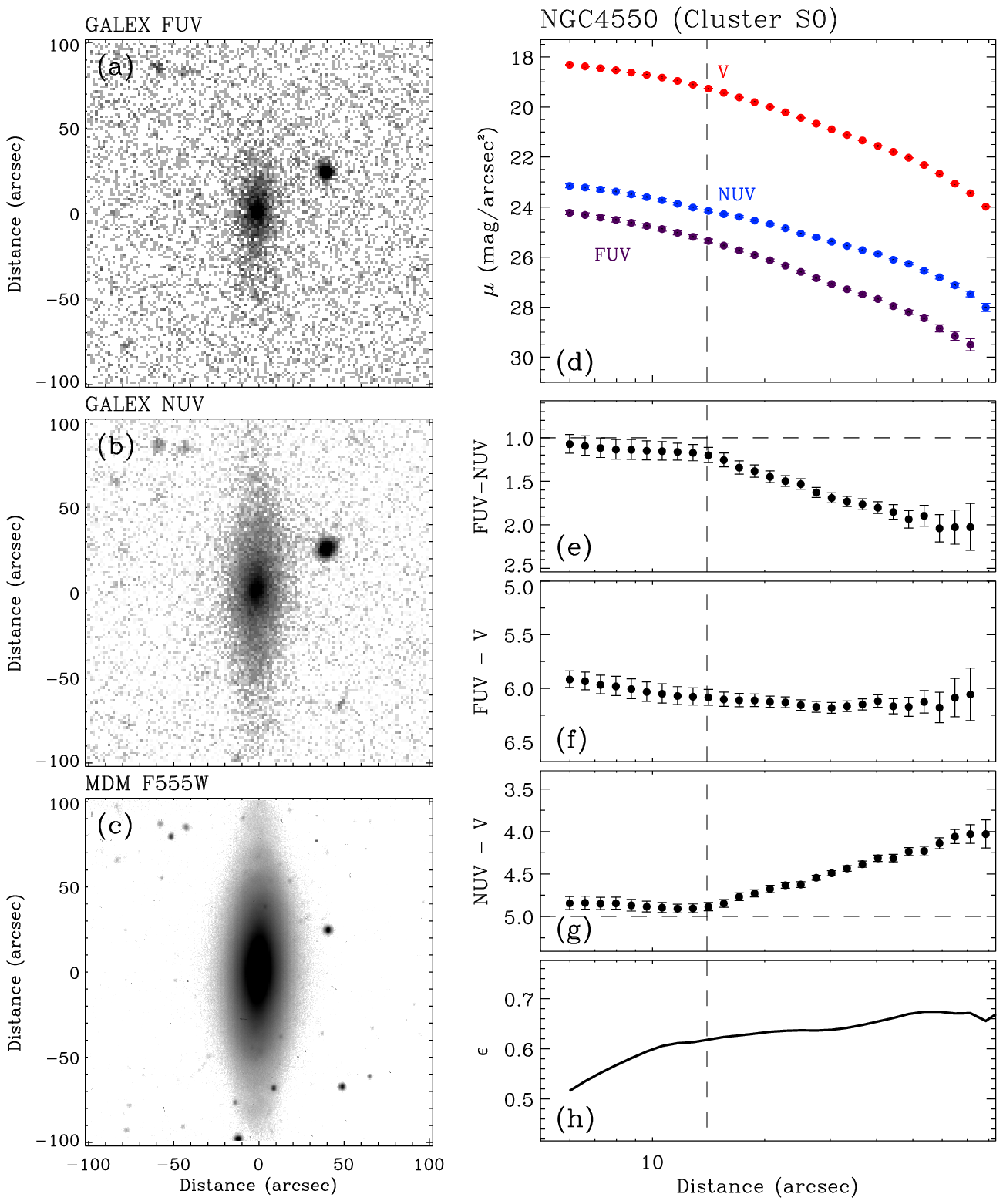}
\vspace*{10mm}
\includegraphics[width=7.7cm]{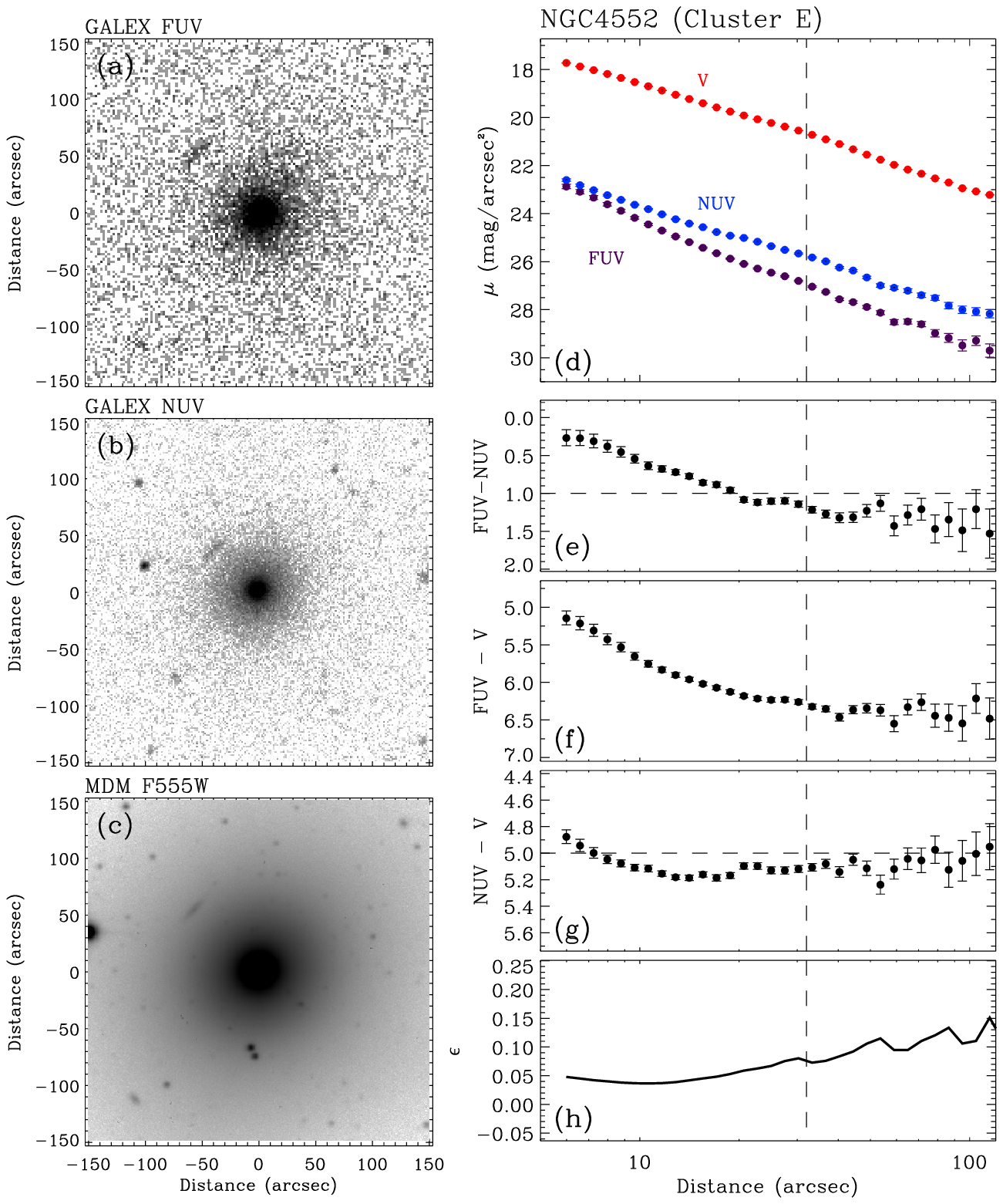}\\
\vspace*{10mm}
\includegraphics[width=7.7cm]{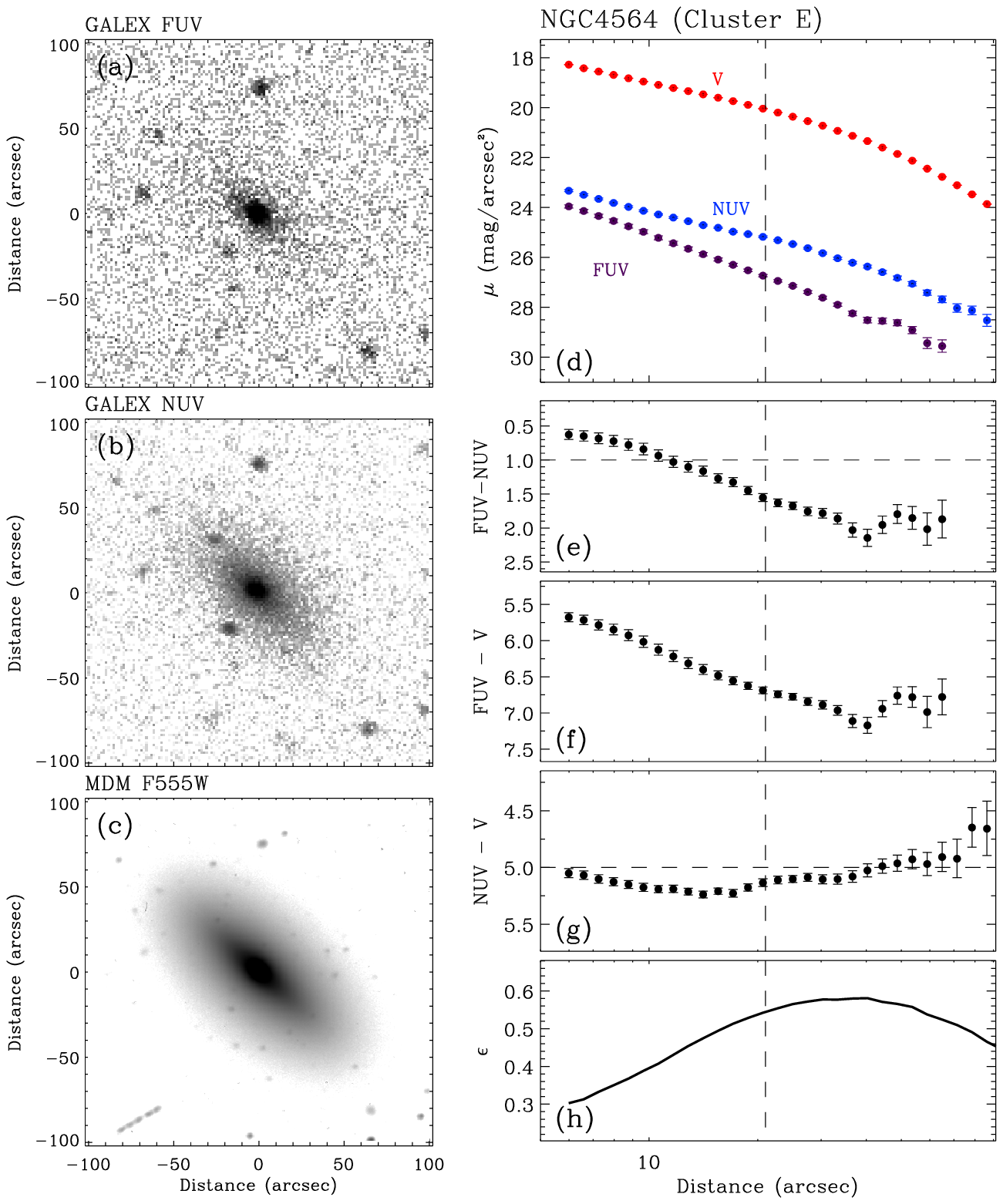}
\vspace*{10mm}
\includegraphics[width=7.7cm]{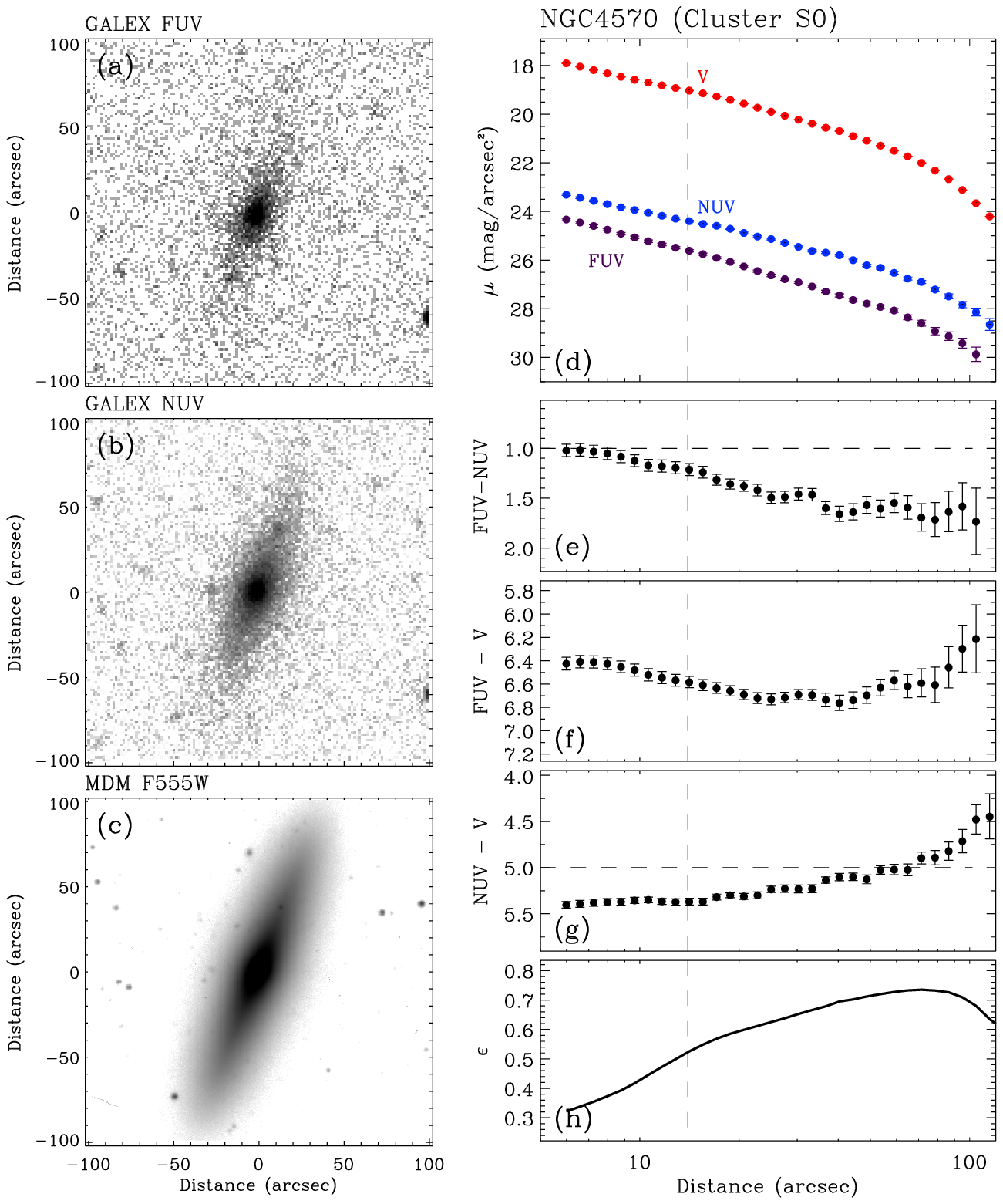}
\end{center}
\caption[]{continued.}
\end{figure*}
\addtocounter{figure}{-1}
\begin{figure*}
\begin{center}
\vspace*{10mm}
\includegraphics[width=7.7cm]{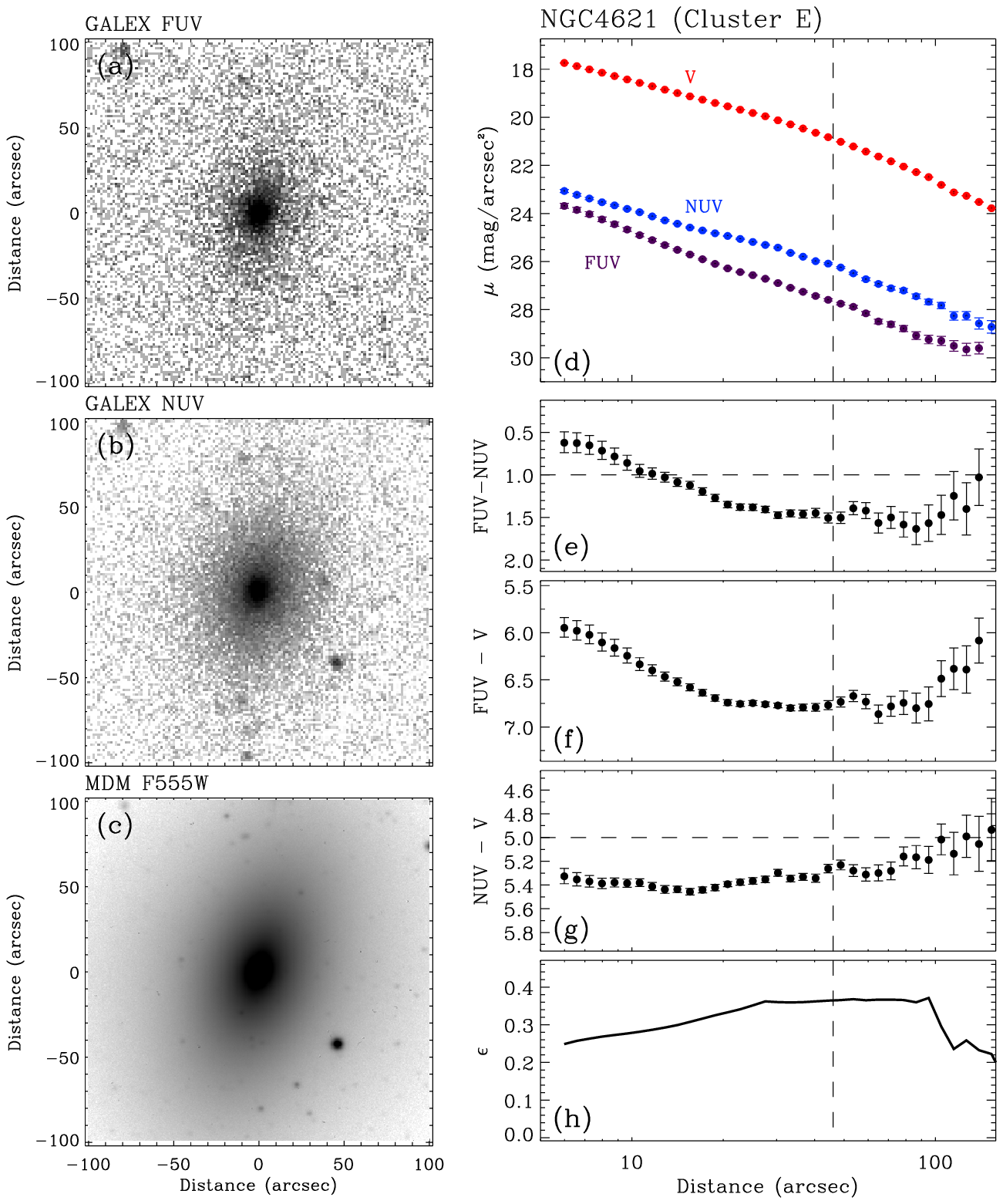}
\vspace*{10mm}
\includegraphics[width=7.7cm]{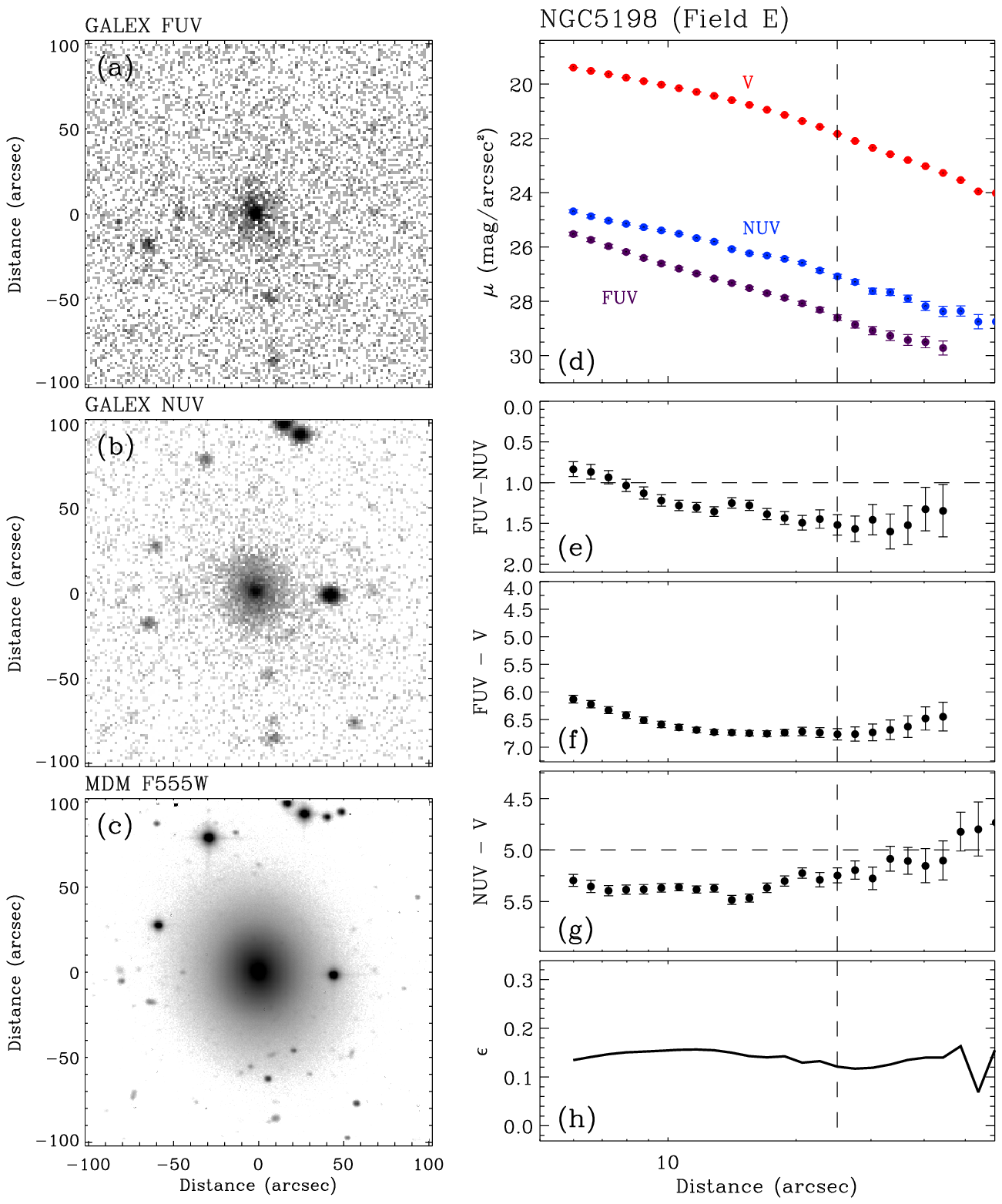}\\
\end{center}
\caption[]{continued.}
\end{figure*}
%
%
%
%
%
%
\subsection{Optical observations}
\label{sec:mdm}
Ground-based optical imaging observations in the {\it Hubble Space
  Telescope} ({\it HST}) filter F555W (similar to Johnson $V$) were
obtained using the MDM Observatory 1.3-m McGraw-Hill Telescope, as
part of a large survey targeting the whole {\tt SAURON} galaxy
sample. The MDM observations are described in detail in
Falc\'{o}n-Barroso et al.\ (in prep.) and were reduced and calibrated
in the standard manner. The field-of-view of the MDM images is
$17\farcm3\times17\farcm3$ with $0\farcs508\times0\farcs508$ pixels,
allowing for accurate sky subtraction and proper sampling of the
seeing. The seeing during the observations was typically
$\approx1.2$~arcsec. The total $V$ magnitudes we use in this paper are
also from Falc\'{o}n-Barroso et al.\ (in prep.).
%
%
\subsection{UV observations}
\label{sec:galex}
We obtained both FUV ($1350$--$1750$~\AA) and NUV ($1750$--$2750$
~\AA) images of $34$ E/S0 galaxies using the medium-depth imaging
mode of {\it GALEX}, as part of an ongoing UV imaging survey of the
{\tt SAURON} sample ({\it GALEX} guest investigator programmes
GI1--109 and GI3--041) and the {\it GALEX} Nearby Galaxy Survey
\citep[NGS;][]{getal07}. Of the $48$ early-type galaxies in the {\tt
SAURON} sample, $6$ are too close to UV-bright stars to be observed
with {\it GALEX} and $3$ from our programmes have not been observed
yet, similarly for $5$ galaxies from other guest investigator
programmes, explaining the current sample of $34$. The typical
exposure time per field was one orbit ($\approx 1700$~s) but
occasionally more. Details of the {\it GALEX} instruments, pipeline
and calibration are described in \citet{maetal05} and
\citet{moetal05,moetal07}. We note in particular a possible
systematic error in the FUV and NUV zero-points of up to $0.15$~mag.
This is not included in the uncertainties quoted in the current
paper as we are mostly interested in the relative colours of our
objects. The spatial resolutions of the images are approximately
$4\farcs5$ and $6\farcs0$ FWHM in FUV and NUV, respectively, sampled
with $1\farcs5\times1\farcs5$ pixels.

%
%
\begin{table*}
\caption{Integrated properties.}
\label{tab:list}
\begin{tabular}{@{}lrrrrrrrrrl}
\hline Galaxy & $E(B\!-\!V)$ & D & $\sigma_{\rm e}$ &
\multicolumn{2}{c}{$R_{\rm e}$} &
\multicolumn{2}{c}{$\langle$$\mu$$\rangle$$_{\rm e}$} &
\multicolumn{2}{c}{Total apparent magnitude} & UV type\\
& & & & FUV & NUV & FUV & NUV & FUV & NUV & \\
& & (Mpc) & (km~s$^{-1}$) & \multicolumn{2}{c}{(arcsec)} & \multicolumn{2}{c}{(mag)} &
\multicolumn{2}{c}{(mag)} & \\
\hline
%
NGC0474 &   0.034   &   26.1   &   150 &   74  &   67  &   28.96   $\pm$   0.49    &   26.96   $\pm$   0.31    &   
17.78   $\pm$   0.37    &   15.82   $\pm$   0.28  &  RSF  \\
NGC0524 &   0.083   &   24.0   &   235 &   17  &   39  &   25.90   $\pm$   0.30    &   25.82   $\pm$   0.26    &   
17.70   $\pm$   0.13    &   15.88   $\pm$   0.15  &  \\
NGC0821 &   0.110   &   24.1   &   189 &   27  &   33  &   27.21   $\pm$   0.50    &   25.84   $\pm$   0.27    &   
18.10   $\pm$   0.36    &   15.77   $\pm$   0.12  & A/Z  \\
NGC1023 &   0.061   &   11.4   &   182 &   44  &   70  &   26.38   $\pm$   0.30    &   25.76   $\pm$   0.25    &   
16.16   $\pm$   0.15    &   14.53   $\pm$   0.12  &  RSF  \\
NGC2695 &   0.018   &   32.4   &   188 &   13  &   27  &   25.89   $\pm$   0.28    &   25.96   $\pm$   0.24    &   
18.32   $\pm$   0.11    &   16.81   $\pm$   0.11  & UVX, A/Z  \\
NGC2699 &   0.020   &   26.9   &   124 &   9   &   27  &   25.93   $\pm$   0.29    &   26.32   $\pm$   0.26    &   
19.06   $\pm$   0.14    &   17.27   $\pm$   0.16  &  A/Z  \\
NGC2768 &   0.044   &   22.4   &   216 &   50  &   77  &   27.86   $\pm$   0.38    &   26.82   $\pm$   0.31    &   
16.75   $\pm$   0.26    &   15.35   $\pm$   0.27  &  \\
NGC2974 &   0.054   &   21.5   &   233 &   11  &   29  &   24.91   $\pm$   0.27    &   25.43   $\pm$   0.23    &   
17.71   $\pm$   0.09    &   16.10   $\pm$   0.09  & RSF  \\
NGC3032 &   0.017   &   22.0   &   90  &   2   &   3   &   20.83   $\pm$   0.25    &   20.43   $\pm$   0.20    &   
16.94   $\pm$   0.05    &   15.78   $\pm$   0.01  &  RSF  \\
NGC4150 &   0.018   &   13.7   &   77  &   5   &   18  &   23.46   $\pm$   0.25    &   24.32   $\pm$   0.21    &   
18.20   $\pm$   0.17    &   16.10   $\pm$   0.03  &  RSF  \\
NGC4278 &   0.029   &   16.1   &   231 &   27  &   44  &   23.63   $\pm$   0.25    &   24.49   $\pm$   0.21    &   
16.41   $\pm$   0.08    &   15.21   $\pm$   0.07  &  UVX, A/Z  \\
NGC4374 &   0.040   &   18.5   &   278 &   38  &   65  &   25.63   $\pm$   0.27    &   25.42   $\pm$   0.23    &   
15.72   $\pm$   0.09    &   14.35   $\pm$   0.08  &  UVX  \\
NGC4387 &   0.033   &   18.0   &   98  &   12  &   15  &   26.33   $\pm$   0.31    &   25.08   $\pm$   0.22    &   
19.03   $\pm$   0.18    &   17.27   $\pm$   0.06  &  \\
NGC4458 &   0.024   &   16.4   &   85  &   13  &   23  &   27.08   $\pm$   0.41    &   25.95   $\pm$   0.25    &   
19.40   $\pm$   0.26    &   16.84   $\pm$   0.11  &  \\
NGC4459 &   0.046   &   16.1   &   168 &   13  &   20  &   24.63   $\pm$   0.26    &   24.13   $\pm$   0.21    &   
17.02   $\pm$   0.07    &   15.49   $\pm$   0.03  &  RSF  \\
NGC4473 &   0.028   &   15.3   &   192 &   25  &   47  &   26.09   $\pm$   0.29    &   25.60   $\pm$   0.23    &   
17.11   $\pm$   0.11    &   15.24   $\pm$   0.08  &  \\
NGC4477 &   0.032   &   16.7   &   162 &   30  &   33  &   27.36   $\pm$   0.40    &   26.04   $\pm$   0.25    &   
17.24   $\pm$   0.15    &   15.48   $\pm$   0.08  &  \\
NGC4486 &   0.022   &   17.2   &   298 &   36  &   51  &   24.29   $\pm$   0.26    &   24.46   $\pm$   0.21    &   
14.50   $\pm$   0.06    &   13.92   $\pm$   0.03  &  UVX  \\
NGC4526 &   0.022   &   16.9   &   222 &   25  &   26  &   25.60   $\pm$   0.27    &   24.24   $\pm$   0.21    &   
16.59   $\pm$   0.09    &   15.22   $\pm$   0.03  &  RSF  \\
NGC4546 &   0.034   &   14.1   &   194 &   23  &   54  &   26.33   $\pm$   0.31    &   25.85   $\pm$   0.24    &   
17.56   $\pm$   0.13    &   15.41   $\pm$   0.13  &  A/Z  \\
NGC4550 &   0.039   &   15.5   &   110 &   7   &   14  &   24.30   $\pm$   0.26    &   24.05   $\pm$   0.21    &   
18.05   $\pm$   0.18    &   16.41   $\pm$   0.03  &  RSF  \\
NGC4552 &   0.041   &   15.9   &   252 &   13  &   19  &   23.39   $\pm$   0.25    &   23.59   $\pm$   0.21    &   
15.85   $\pm$   0.06    &   14.94   $\pm$   0.02  &  UVX  \\
NGC4564 &   0.035   &   15.9   &   155 &   5   &   21  &   23.15   $\pm$   0.25    &   24.65   $\pm$   0.21    &   
17.71   $\pm$   0.06    &   16.09   $\pm$   0.04  &  UVX, A/Z  \\
NGC4570 &   0.022   &   17.1   &   173 &   13  &   26  &   25.00   $\pm$   0.26    &   25.00   $\pm$   0.22    &   
17.52   $\pm$   0.08    &   15.95   $\pm$   0.05  &  A/Z \\
NGC4621 &   0.033   &   14.9   &   211 &   11  &   35  &   23.83   $\pm$   0.26    &   24.93   $\pm$   0.22    &   
16.65   $\pm$   0.06    &   15.25   $\pm$   0.05  &  UVX  \\
NGC5198 &   0.023   &   38.4   &   179 &   7   &   18  &   24.92   $\pm$   0.27    &   25.47   $\pm$   0.23    &   
18.86   $\pm$   0.11    &   17.23   $\pm$   0.08  &  UVX, A/Z  \\
NGC5308 &   0.018   &   29.4   &   208 &   12  &   25  &   25.75   $\pm$   0.28    &   25.66   $\pm$   0.23    &   
18.36   $\pm$   0.11    &   16.81   $\pm$   0.09  &  A/Z  \\
NGC5813 &   0.057   &   32.2   &   230 &   40  &   80  &   27.20   $\pm$   0.35    &   26.82   $\pm$   0.32    &   
17.19   $\pm$   0.24    &   15.50   $\pm$   0.34  &  A/Z  \\
NGC5831 &   0.059   &   27.2   &   151 &   15  &   55  &   26.23   $\pm$   0.30    &   26.70   $\pm$   0.31    &   
18.36   $\pm$   0.15    &   16.20   $\pm$   0.32  &  A/Z  \\
NGC5838 &   0.053   &   24.7   &   240 &   10  &   21  &   24.81   $\pm$   0.27    &   24.73   $\pm$   0.22    &   
17.77   $\pm$   0.08    &   16.16   $\pm$   0.05  &  \\
NGC5845 &   0.053   &   25.9   &   239 &   6   &   6   &   24.59   $\pm$   0.27    &   23.59   $\pm$   0.21    &   
18.78   $\pm$   0.12    &   17.70   $\pm$   0.02  &  A/Z  \\
NGC5846 &   0.055   &   24.9   &   238 &   22  &   32  &   25.37   $\pm$   0.27    &   25.15   $\pm$   0.23    &   
16.66   $\pm$   0.09    &   15.51   $\pm$   0.07  &  UVX  \\
NGC5982 &   0.018   &   51.7   &   229 &   12  &   29  &   25.19   $\pm$   0.27    &   25.58   $\pm$   0.23    &   
17.76   $\pm$   0.08    &   16.25   $\pm$   0.08  &  UVX, A/Z  \\
NGC7457 &   0.052   &   13.2   &   78  &   37  &   43  &   26.35   $\pm$   0.30    &   25.89   $\pm$   0.25    &   
18.39   $\pm$   0.31    &   15.83   $\pm$   0.13  &  RSF  \\
\hline
\end{tabular}
Columns: (1) Galaxy identifier. (2) Galactic extinction in the
NASA/IPAC Extragalactic Database (NED) from \citet*{schetal98}. (3)
Distance from Falc\'{o}n-Barroso et al.\ (in prep.). (4) Velocity
dispersion of the luminosity-weighted spectrum within a circular
aperture of $R_{\rm e}$ from \citeauthor{eetal07}. The uncertainty
is taken as $5$~per cent. (5)--(6) Effective radius in each band.
The uncertainties are taken as $20$~per cent. (7)--(8) $Mean$
effective surface brightness in each band and error. (9)--(10) Total
apparent magnitude in each band and error. (11) UV$-$optical radial
colour profiles classification. RSF: recent star formation. UVX:
UV-upturn. A/Z: large-scale age and/or metallicity gradient.
\end{table*}
\subsection{Photometry}
\label{sec:results}
The {\it GALEX} UV images are delivered pre-processed but we
undertook our own estimate of the sky values. This was calculated
for each image as the mean of the sky values in $90$--$100$ image
regions after masking-out Sextractor-detected sources. Typical mean
sky values were $0.4$~counts~pixel$^{-1}$ in FUV and
$4$~counts~pixel$^{-1}$ in NUV. We computed the mean sky value
uncertainty in each image from the distribution of sky values in the
small regions assuming a Gaussian distritubion. These were typically
$0.06$~counts~pixel$^{-1}$ in FUV and $0.2$~counts~pixel$^{-1}$ in
NUV. We took the low level flat fielding errors to be negligible,
so these measurements form the basis of the errors plotted in the
individual FUV and NUV radial surface brightness profiles of
Figure~\ref{fig:results}. At large radii, the galaxies are faint
and only a few counts are detected in each pixel. Accurately
determining the level of the sky, even if faint, is thus the main
factor limiting the depth of the images and our surface photometry.
The UV radial profiles in Figure~\ref{fig:results} are truncated
when the uncertainty in the surface brightness exceeds
$0.3$~mag~arcsec$^{-2}$. The V band profiles are truncated at
the same radial distance where their corresponding NUV profiles end.

We convolved the FUV and optical data to the spatial resolution of
the NUV observations before any analysis, to avoid spurious colour
gradients in the inner parts. We carried out surface photometry by
measuring the surface brightness along elliptical annuli, using the
iterative method described by \citet{j87} and implemented in the
{\small ELLIPSE} task within the {\small STSDAS ISOPHOTE} package in
{\small IRAF} (Image Reduction and Analysis Facility). The centre of
the isophotes was fixed to the centre of the light distribution and
the position angle (PA), ellipticity ($\epsilon$) and surface
brightness ($\mu$) were fitted as a function of the radius. Ellipses
were fitted to the $V$-band images only, which have a far superior
signal-to-noise (S/N) ratio at all radii compared with the UV
images, and were then imposed on the UV images so that meaningful
colours could be derived. Galactic extinction was corrected with
$R_V=3.1$ \citep*{cetal89}, $A_{\rm FUV}=8.376\times\,E(B-V)$ and
$A_{\rm NUV}=8.741\times\,E(B-V)$ \citep{wetal05} using reddening
maps from \citet*{schetal98}. We present the {\it GALEX} FUV, NUV
and MDM F555W images as well as the surface-brightness, colour and
ellipticity radial profiles in Figure~\ref{fig:results}. The
vertical dashed lines in Figure~\ref{fig:results} show the effective
radii determined by the {\tt SAURON} survey at $I$ band (see
\citeauthor{ketal06}). For reference, a UV spectral slope of zero in
the $\lambda$-$F_\lambda$ plane is roughly the same as
FUV$-$NUV$=1.0$ in AB magnitudes (shown as a horizontal line in the
FUV$-$NUV colour profiles of Figure~\ref{fig:results}). We give
comments on notable features in the profiles of individual galaxies
in Appendix~\ref{sec:list}.

%
%
\begin{figure*}
\begin{center}
\includegraphics[width=13cm]{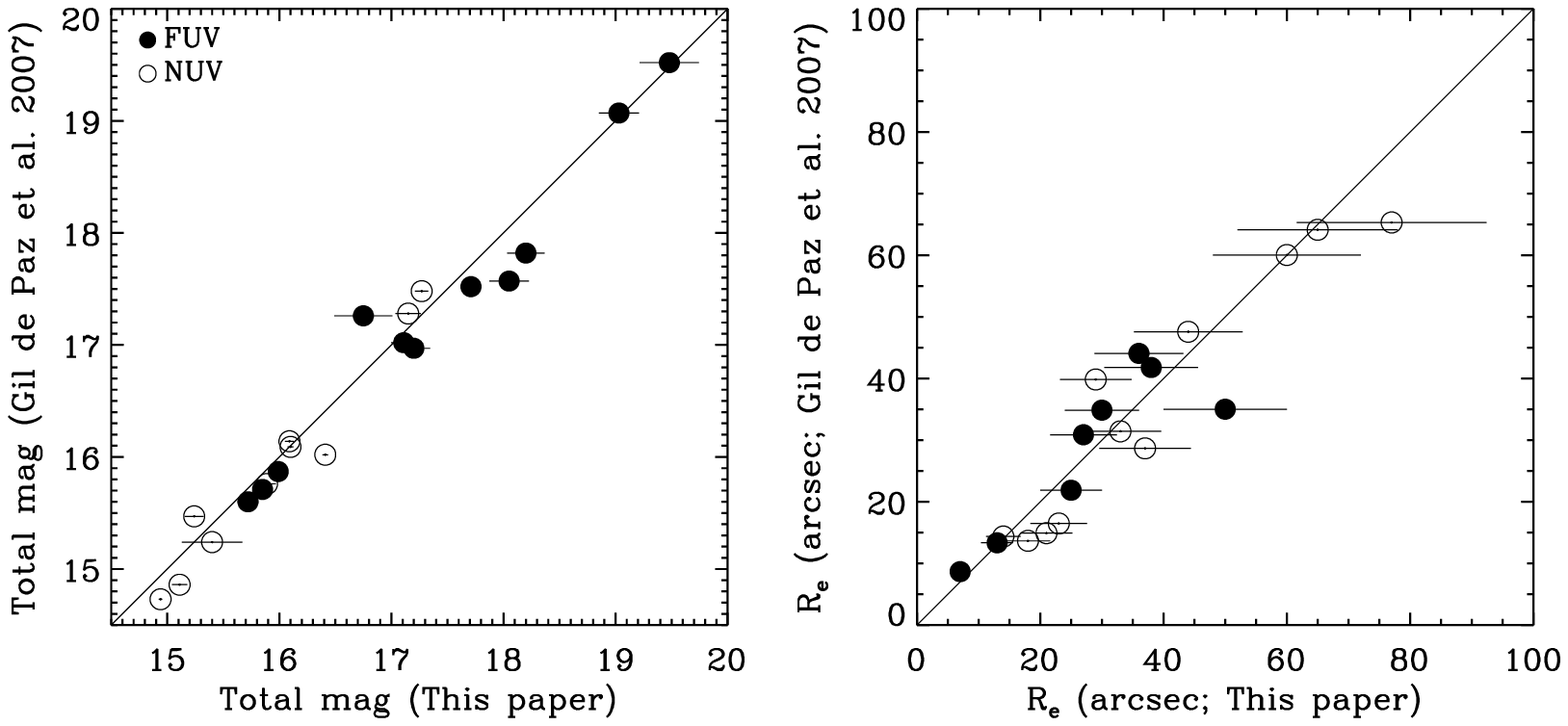}
\end{center}
\caption[]{Comparison of our magnitudes and effective radii in the $FUV$ and $NUV$
passbands with those of \citet{getal07}.}
\label{fig:com}
\end{figure*}

From the radial surface brightness profiles, we derived total apparent
magnitudes by extrapolating the growth curves to infinity. For each
profile, we first calculated the integrated apparent magnitude
($m(R)$) within elliptical isophotes up to the radius where the
uncertainty in the surface brightness reaches
$0.3$~mag~arcsec$^{-2}$. We then computed the slope of the growth
curve ($S\equiv dm(R)/dR$) within that region, plotted it as a
function of the integrated magnitude $m(R)$, and fitted a straight line
to the outer parts. The value of $m$ where the extrapolated value of
$S$ is $0$ was taken as the total apparent magnitude. Selecting a
suitable radial range for growth curve fitting was occasionally
difficult in FUV because of the low S/N in the outer parts, but the
errors quoted correspond to the uncertainties associated with the fit
to the (slope of the) growth curve only, so they are probably slightly
underestimated. This technique is described in more detail in
\citet{cetal01} and \citet{getal07}. The effective radius ($R_{\rm
  e}$; semi-major axis of the elliptical aperture containing half the
light), the mean surface brightness within $R_{\rm e}$
($\langle\mu\rangle_{\rm e}$) and their uncertainties were also
calculated from the growth curve. Having said that, the uncertainty in
$R_{\rm e}$ arising from different measurement methods is always
larger than formal uncertainties, and a comparison of different
measurements shows that it is typically $\approx20$~per cent, which we
adopt (see Falc\'{o}n-Barroso et al., in prep.).

In Figure~\ref{fig:com}, we show the comparison of our magnitudes and
effective radii in both FUV and NUV with those of \citet{getal07}. The
agreement is generally good. The mean of the absolute values of the
total apparent magnitude differences is $\approx0.16$ mag in both FUV
and NUV, as expected generally somewhat larger than our formal
uncertainties. For the effective radii, the mean absolute difference
is $\approx9$~per cent in FUV and $\approx15$~per cent in NUV, so
generally within our errors. The integrated UV and other properties
are listed in Table~\ref{tab:list}.

%
%
\section{DISCUSSION}
\label{sec:discussion}
%
%
\subsection{UV colour-magnitude relations}
\label{CMRs}
%
%
\begin{figure}
\begin{center}
\includegraphics[width=8cm]{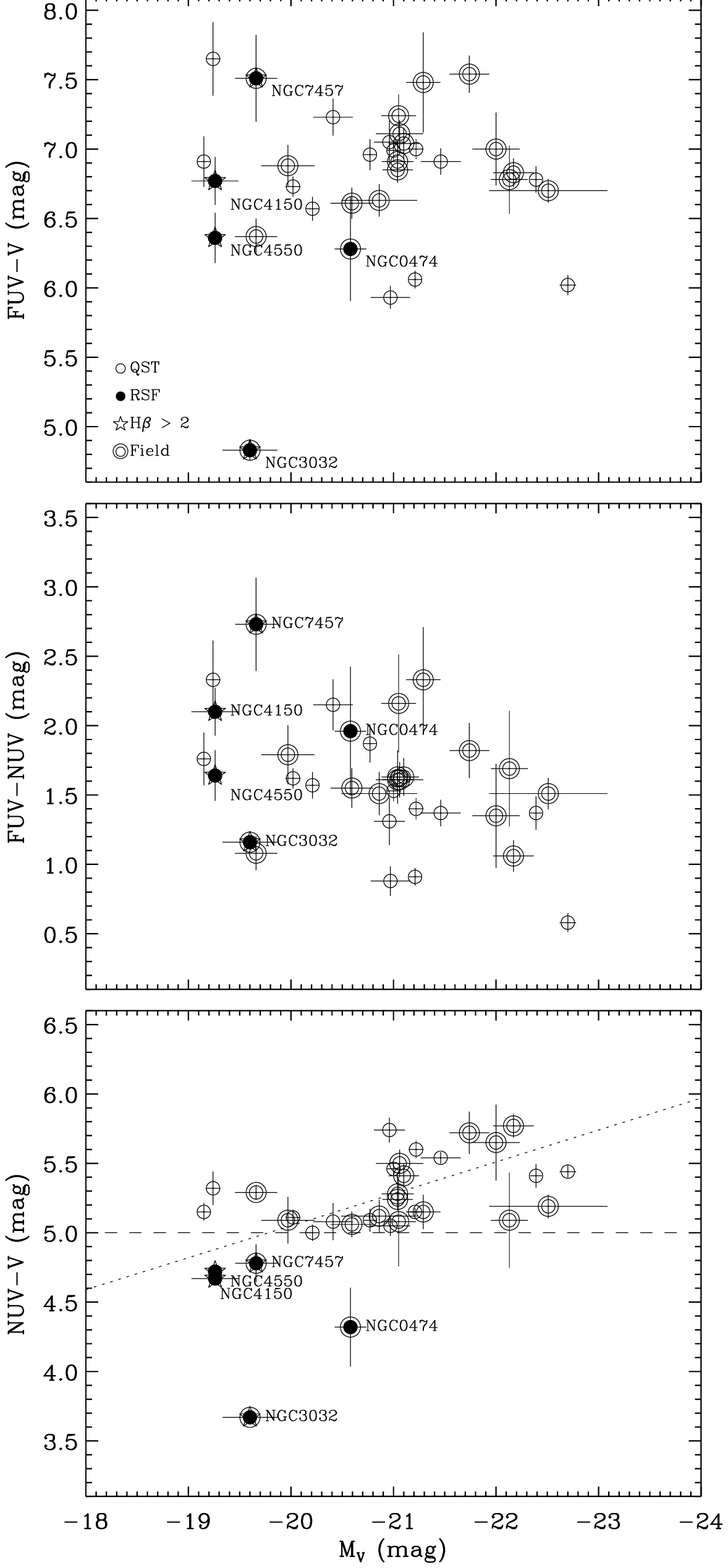}
\end{center}
\caption[]{UV colour-magnitude relations of our $34$ {\tt SAURON}
early-type galaxies. The FUV$-V$ ({\em top}), FUV$-$NUV ({\em
middle}) and NUV$-V$ ({\em bottom}) colours are shown as a function
of the absolute magnitude in $V$. The dashed line indicates the
NUV$-V=5.0$ cutoff for recent star formation. Using this standard,
we divided the sample into quiescent early-type galaxies (QST, open
symbols) and recent star formation (RSF, filled symbols) galaxies.
Stars denote galaxies with H$\beta\ga2$~\AA. Field galaxies are
represented by two concentric circles. The red-sequence fit from
\citet{yetal05} in NUV$-r$ is shown as a dotted line in the NUV$-V$
colour-magnitude relation (with offset applied, see text).}
\label{fig:CMRs}
\end{figure}

The simplest new data product obtained from our {\it GALEX}
observations is the integrated UV apparent magnitudes, from which we
derive total absolute magnitudes using the distances listed in
Table~\ref{tab:list}. Colour-magnitude relations are widely used to
study the star formation history (SFH) of early-type galaxies. The NUV
CMR \citep[see, e.g.,][]{yetal05, ketal07} is a particularly good tool
for tracking recent star formation, owing to its high sensitivity to
the presence of young stellar populations. Figure~\ref{fig:CMRs} shows
the UV--optical CMRs. It also shows the red sequence NUV$-r$ fit of
\citet{yetal05}, after correcting for the sky measurement offset from
the {\it GALEX} pipeline \citep[see][]{getal07}. The scatter in the
CMRs of Figure~\ref{fig:CMRs} is significant and there are clearly
outliers about the red sequence relation. We henceforth attempt to
quantify the effects of recent star formation on those relations.

Most UV photons are emitted by stars younger than a few hundred
megayears, so they are useful to trace residual star formation, but
FUV flux can also be generated by evolved hot helium-burning stars
(the so-called UV-upturn phenomenon; see, e.g., \citealt*{ydo97} and
\citealt{oco99} for reviews). Star formation studies thus usually use
the NUV passband instead \citep{yetal05}, although it is nevertheless
important to know how much of the NUV flux can originate from the
UV-upturn phenomenon. NGC~4552 is a famous UV-upturn galaxy; its
FUV$-V$ colour is one of the bluest among nearby elliptical galaxies
with no sign of recent star formation. We measure a NUV$-V$ colour of
$5.15\pm0.03$~mag for NGC~4552, so considering the measurement error
we adopt a very conservative $5\sigma$ upper limit of NUV$-V\,=\,5.0$
to pick out galaxies that have experienced recent star formation.
This empirical demarcation for identifying recent star formation
galaxies is shown as a horizontal dashed line in the NUV$-V$ colour
profiles of Figure~\ref{fig:results} and the NUV$-V$ CMR of
Figure~\ref{fig:CMRs}. We henceforth assume that the relative FUV
strength of NGC~4552 is the maximum that can be generated by purely
old-star populations. This is an {\it ad hoc} assumption, however it
is frequently made and is empirically grounded.

Using this empirical criterion, we have identified the galaxies that
are likely to have experienced a recent episode of star formation in
Figure~\ref{fig:CMRs} onward (filled symbols). We label galaxies with
NUV$-V\,\ge\,5.0$ as quiescent early-type galaxies (QST, open
symbols). Furthermore, stars and concentric circles indicate Balmer
absorption line strengths H$\beta\ga\,2$~\AA\ and field galaxies,
respectively. The H$\beta$ indices are computed within one effective
radius in the optical and come from the {\tt SAURON} data published in
\citeauthor{ketal06}. Supporting our empirical threshold, four out of
five recent star formation candidate galaxies (NGC~3032, NGC~4150,
NGC~4550 and NGC~7457) also show enhanced H$\beta$ line strengths, a
widely-used post-starburst signature. The only exception (NGC~474) is
noted for its shell structures located in the outer regions, at
$R\approx60$~arcsec, and as shown in Figure~\ref{fig:results} the
total NUV flux is dominated by these outer regions, whereas the Balmer
line was measured within $R_{\rm e}$ in the optical.

A more in-depth look at Figure~\ref{fig:CMRs}, with likely recent star
formation galaxies identified, reveals that a non-negligible fraction
of the scatter in the UV CMRs (and the departure from the
\citealt{yetal05} red sequence relation) is due to galaxies with
recent star formation. For example, the scatter in the (NUV$-V$)-M$_V$
CMR increases by $67$~per cent between least-square fits excluding and
including RSF candidates. The overall fraction of galaxies with RSF is
$15$~per cent ($5/34$). This may appear to also agree with the result
of \citet{yetal05}, but a direct comparison is not sensible because
the sampling strategies are radically different. It should also be
noted that no clear dependence of FUV$-V$ or FUV$-$NUV on $M_V$ is
visible in our sample. Finally, environment does not appear to play a
significant role in star formation, as the RSF galaxies selected
contain both field and cluster galaxies. We recall however that the
densest environment probed by the {\tt SAURON} survey is that of the
Virgo cluster of galaxies (see \citeauthor{zetal02}), and that our
sample is admittedly too small to robustly investigate environmental
effects.

%
%
\begin{figure*}
\begin{center}
\includegraphics[width=8cm]{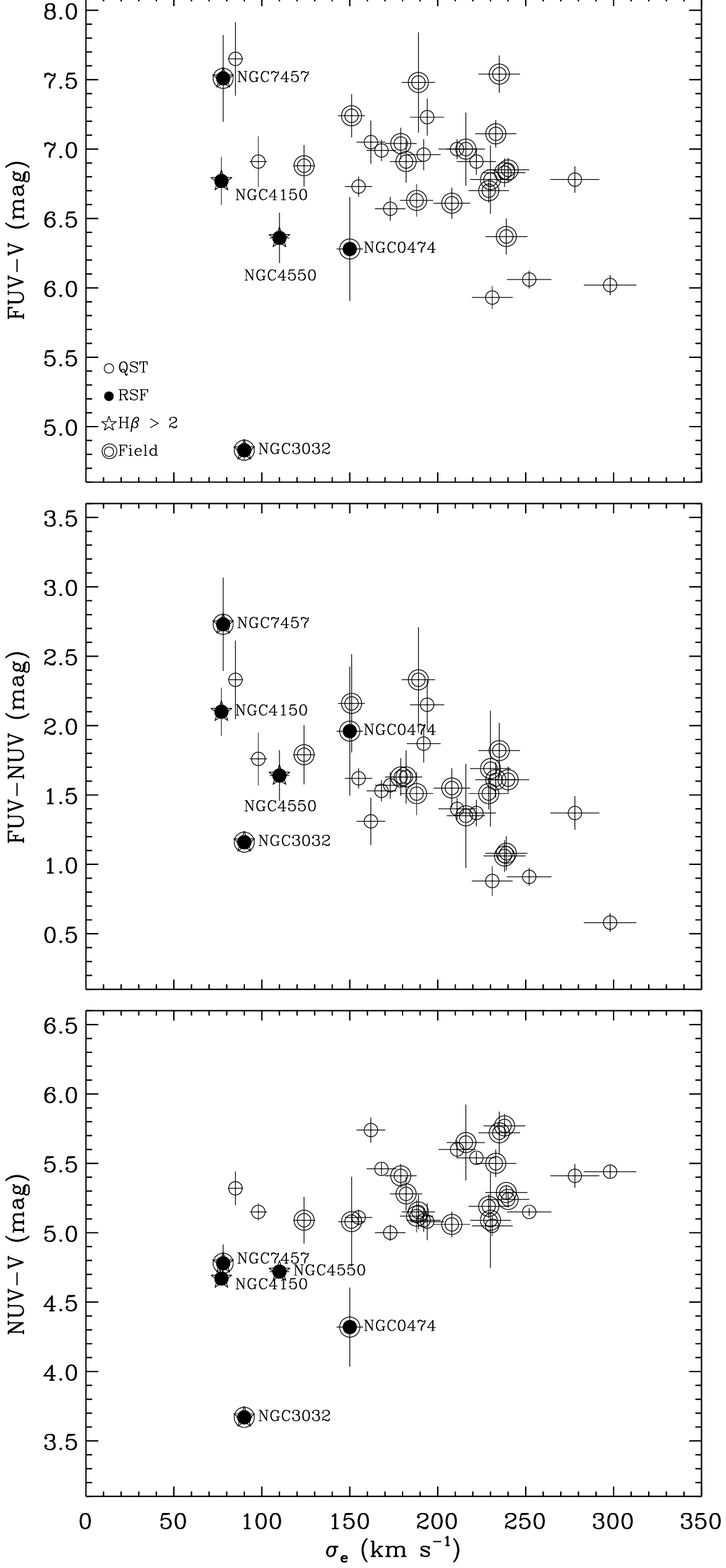}
\includegraphics[width=8cm]{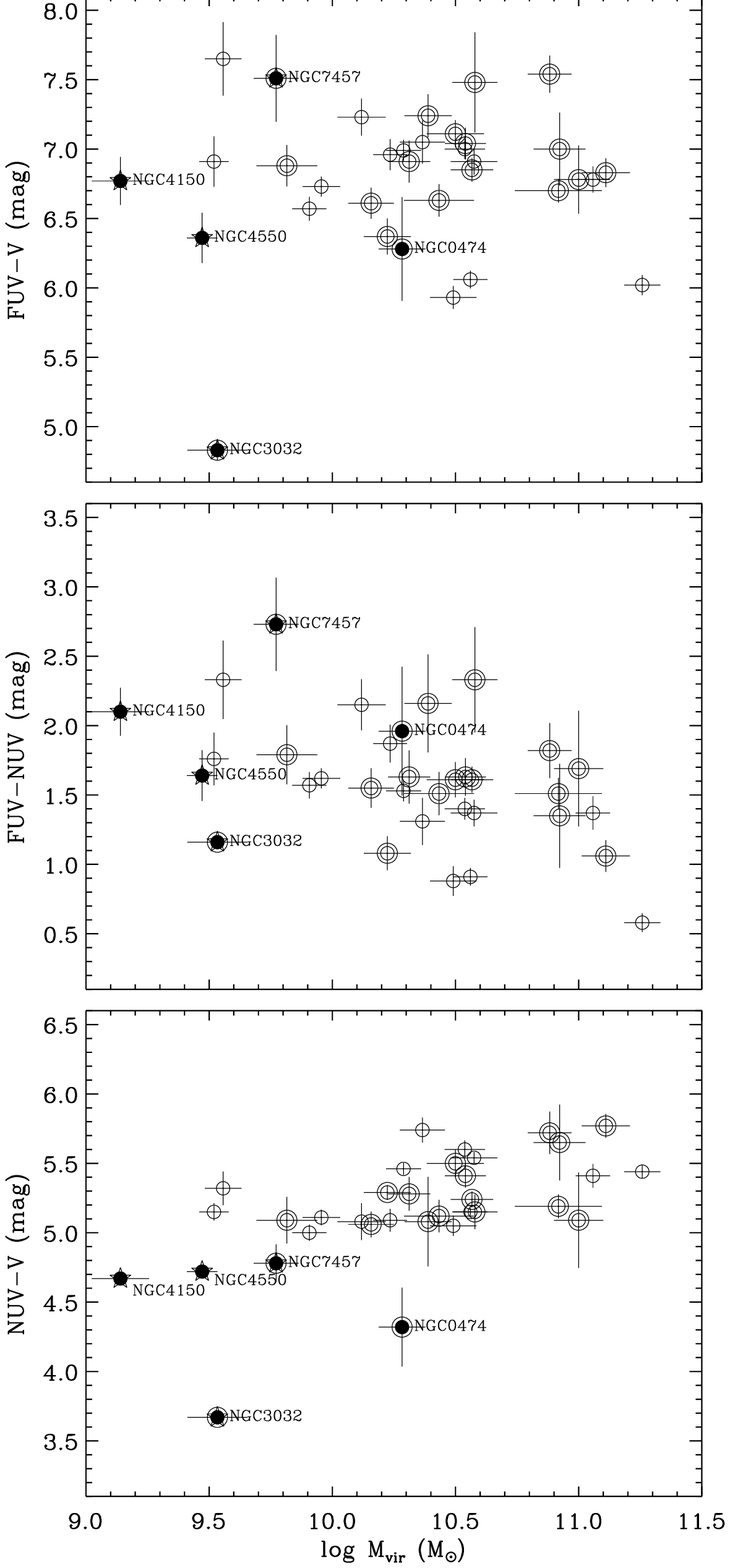}\\
\end{center}
\caption[]{Colour-stellar velocity dispersion and colour-mass
  relations of our $34$ {\tt SAURON} early-type galaxies. Symbols are
  the same as in Figure~\ref{fig:CMRs}.}
\label{fig:CMRs2}
\end{figure*}

We also plot the colour-$\sigma_{\rm e}$ (integrated stellar
velocity dispersion within $R_{\rm e}$) and colour-mass (virial mass
estimate) relations in Figure~\ref{fig:CMRs2}. The dynamical mass
estimates are of the form $M_{\rm vir}\,=\,(5.0\pm0.1)\times R_{\rm
e}\,\sigma_{\rm e}^2/G$ (\citeauthor{cetal06}). Recent star
formation galaxies generally have lower velocity dispersions and
correspondingly smaller virial masses than the majority of red
early-type galaxies, and there is no RSF early-type galaxy with a
velocity dispersion $\sigma_{\rm e}\,>\,200$~km~s$^{-1}$ \citep[see
also][]{scetal06}. This implies that only low-mass early-type
galaxies have formed stars recently, confirming the results of
\citet{scetal07}. In addition, the more massive (high-velocity
dispersion) early-type galaxies tend to be bluer in {\em both}
FUV$-V$ and FUV$-$NUV, confirming earlier studies
\citep[e.g.][]{betal88,detal07}.

Interestingly, using {\it Spitzer} Infrared Array Camera (IRAC)
imaging and Infrared Spectrograph (IRS) spectroscopy, Shapiro et
al.\ (in prep.) surveyed the mechanisms driving star formation in
the {\tt SAURON} early-type galaxies, and found that star formation
happens exclusively in fast-rotating systems (see
\citeauthor{eetal07}). All of our recent star formation galaxies are
also classified as star-forming galaxies in their survey, except
again for NGC~474, and all are fast-rotating except NGC~4550, the peculiar
galaxy with co-spatial counter-rotating discs (extensively discussed in
\citeauthor{eetal07}).

%
%
\begin{table*}
  \caption{CO detection rates in recent star formation and quiescent
    early-type galaxies.} \label{tab:co}
\begin{tabular}{@{}lcccc}
\hline
& RSF & Partial RSF & QST\\
\hline
Number of sample galaxies & $5$ & $4$ & $25$\\
Galaxies detected in CO & $3$ & $2$ & $4$\\
Rate (percent) & $60\pm22$ & $50\pm25$ & $16\pm7$\\
\hline
Note: Errors assume a binomial distribution.
\end{tabular}
\end{table*}

Our star formation interpretation is also consistent with the results
of surveys of the molecular gas emission in early-type galaxies by
\citet*{fc07}, \citet{croetal08} and \citet*{ybc08}. CO emission is
detected in three of the RSF candidates (NGC~3032, NGC~4150 and
NGC~4550). In the case of NGC~7457, a molecular gas mass of
$4.4\,\times\,10^6$~$M_\odot$ was reported by \citet{ws03}, but
\citet{fc07} later failed to detect it. NGC~474 was also not detected
by \citet{fc07}, but the $24$~arcsec primary beam did not encompass
the shells and strong NUV emission. We summarize the CO detection
rates of our sample galaxies in Table~\ref{tab:co}, where we have
divided the sample into three categories: RSF, partial RSF and
quiescent galaxies. Again, RSF are galaxies selected by the empirical
criterion NUV$-V\,<\,5.0$. Partial RSF galaxies do not satisfy the
overall blue colour criterion in integrated NUV$-V$, but they show
significant blue regions in their radial profiles. Quiescent galaxies
have NUV$-V\,\ge\,5.0$ at essentially all radii. The CO detection rate
decreases monotonically from RSF to quiescent galaxies, thus
supporting our RSF identifications and suggesting, unsurprisingly,
that CO-rich galaxies tend to have had more recent star formation.
%
%
\subsection{UV Fundamental Planes}
\label{sec:FP}
%
%
\begin{figure}
\begin{center}
\includegraphics[width=8cm]{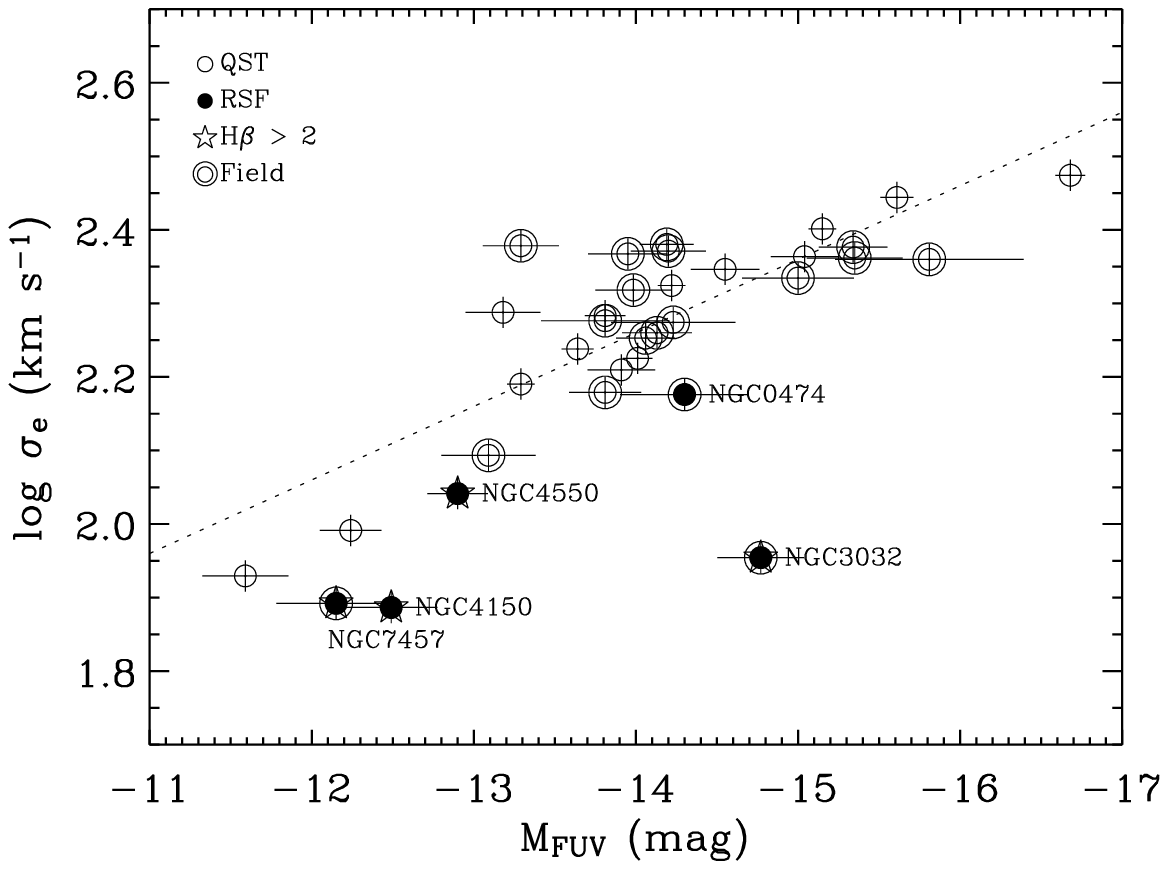}
\includegraphics[width=8cm]{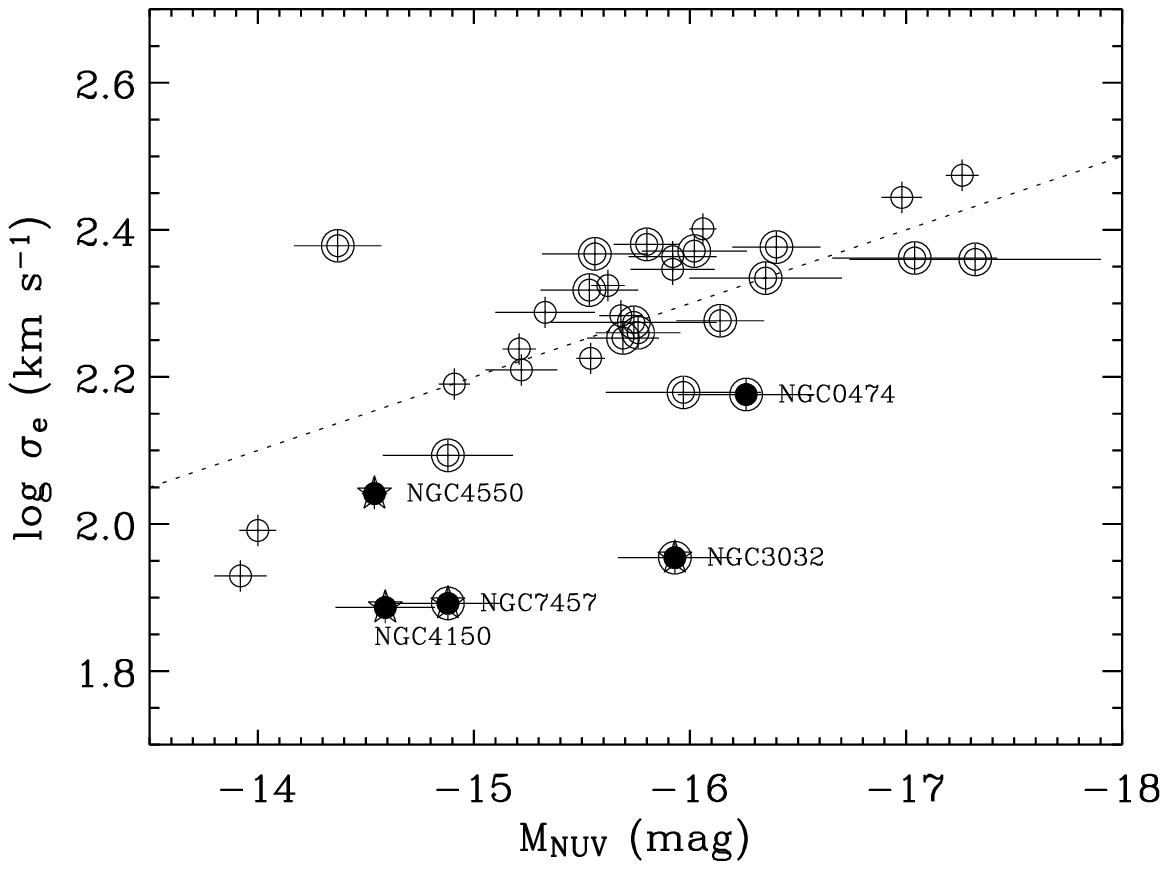}\\
\end{center}
\caption[]{Faber-Jackson relation for our sample of $34$ {\tt SAURON}
  early-type galaxies in the FUV ({\em top}) and NUV ({\em bottom})
  passbands. Symbols are the same as in Figure~\ref{fig:CMRs}. The
  original Faber-Jackson relation ($\sigma\,\propto\,L^{1/4}$) is
  shown as a dotted line in each panel.}
\label{fig:FJ}
\end{figure}

Early-type galaxies follow a correlation between total luminosity and
central velocity dispersion of the form $L\,\propto\,\sigma^{q}$,
where $q\approx4$ with considerable scatter. We present this
Faber-Jackson relation \citep{fj76} for our sample galaxies in
Figure~\ref{fig:FJ} for both FUV and NUV bands, using the same symbols
as in Figure~\ref{fig:CMRs} and with a line of slope $4$ overlayed. It
is striking that the galaxies previously identified as RSF are among
the greatest outliers in both relations, where they are responsible
for much of the total scatter. When RSF early-types are included, the
scatter increases by $65$ and $69$~per cent in the FUV and NUV,
respectively, compared to that when only quiescent galaxies are
considered\footnote{For reference, simple linear fits to the whole
sample yield the slopes of
$3.59^{+0.64}_{-0.47}$ (FUV) and $3.09^{+0.73}_{-0.50}$ (NUV), whereas
for the quiescent sub-sample we find $4.12^{+0.60}_{-0.46}$ (FUV) and
$3.63^{+0.74}_{-0.52}$ (NUV).}.
This is consistent with the conclusions from our analysis
of the colour-magnitude relations in the previous sub-section.

We now consider the Fundamental Plane \citep{dd87,detal87}, a relation
between the photometric effective radius ($R_{\rm e}$), the mean
surface brightness within $R_{\rm e}$ ($\langle\mu\rangle_{\rm e}$)
and the effective stellar velocity dispersion ($\sigma_{\rm e}$) of
the form
\begin{equation}
\log(R_{\rm e})\,=\,a\,\log(\sigma_{\rm
e})\,+\,b\log(\langle\mu\rangle_{\rm e})\,+\,c\,\,,
\end{equation}
where $a$, $b$ and $c$ are constant coefficients derived by minimizing
the residuals from the plane, and $a=2.0$ and $b=0.4$ are expected
from the virial theorem if the galaxies form a homologous family with
a constant mass-to-light ratio. Just as for Faber-Jackson, our
velocity dispersions are measured in the optical (roughly $V$ band),
while our photometric quantities are measured in the UV. The
Fundamental Plane of
course implicitly relies on all quantities used in the virial theorem
being measured for the same stellar population. Our UV Fundamental Planes
are thus
somewhat hybrid, and one should keep in mind that this may introduce
some biases (because of both differing apertures and
wavebands). However, as UV-derived stellar kinematics is still some
time away, this approach is necessary and nevertheless illuminating.

The coefficients $a$ and $b$ measured by \citet{jetal96} are
$1.24\pm0.07$ and $0.33\pm0.02$ in the Gunn $r$-band, respectively.
Early investigations suggested that a variation of the coefficient $a$
in different filters would be expected if there were a systematic
variation of the mass-to-light ratio of the stellar populations with
galaxy luminosity or mass \citep[e.g.][]{ps96}. \citet*{petal98b}
indeed reported an increase of the slope with increasing wavelength
(from $U$ to $K$), although \citet{beetal03} found that the FP
coefficients were approximately the same in $g$, $r$, $i$ and $z$
bands using a much larger but more distant sample of early-type
galaxies from SDSS. Meanwhile, \citet*{tretal04} suggested that the FP
tilt is mostly driven by non-homology and that stellar population
effects account for only a small fraction of it. Early-types in field
environments are also generally more diverse and are found to show a
greater scatter in the Fundamental Plane \citep[see, e.g.,][]{dd92,zw93}. These
authors thus suggested that environment plays an important role in the
process of galaxy formation and evolution. Recently, new approaches
have aimed to elucidate the FP tilt, for example via gravitational
lensing \citep{tretal06,boetal08} and stellar dynamical modeling
(\citeauthor{cetal06}). These studies differ from previous ones in
that they do not depend on simple virial assumptions, but measure
accurate dynamical masses directly. Using two different approaches,
they consistently conclude that the tilt is almost entirely due to a
genuine mass-to-light ratio variation and not to
non-homology. However, it still remains unclear just how much of this
variation can be attributed to a change in the dark matter fraction or
to differences in the stellar populations.

%
%
\begin{figure*}
\begin{center}
\includegraphics[width=14cm]{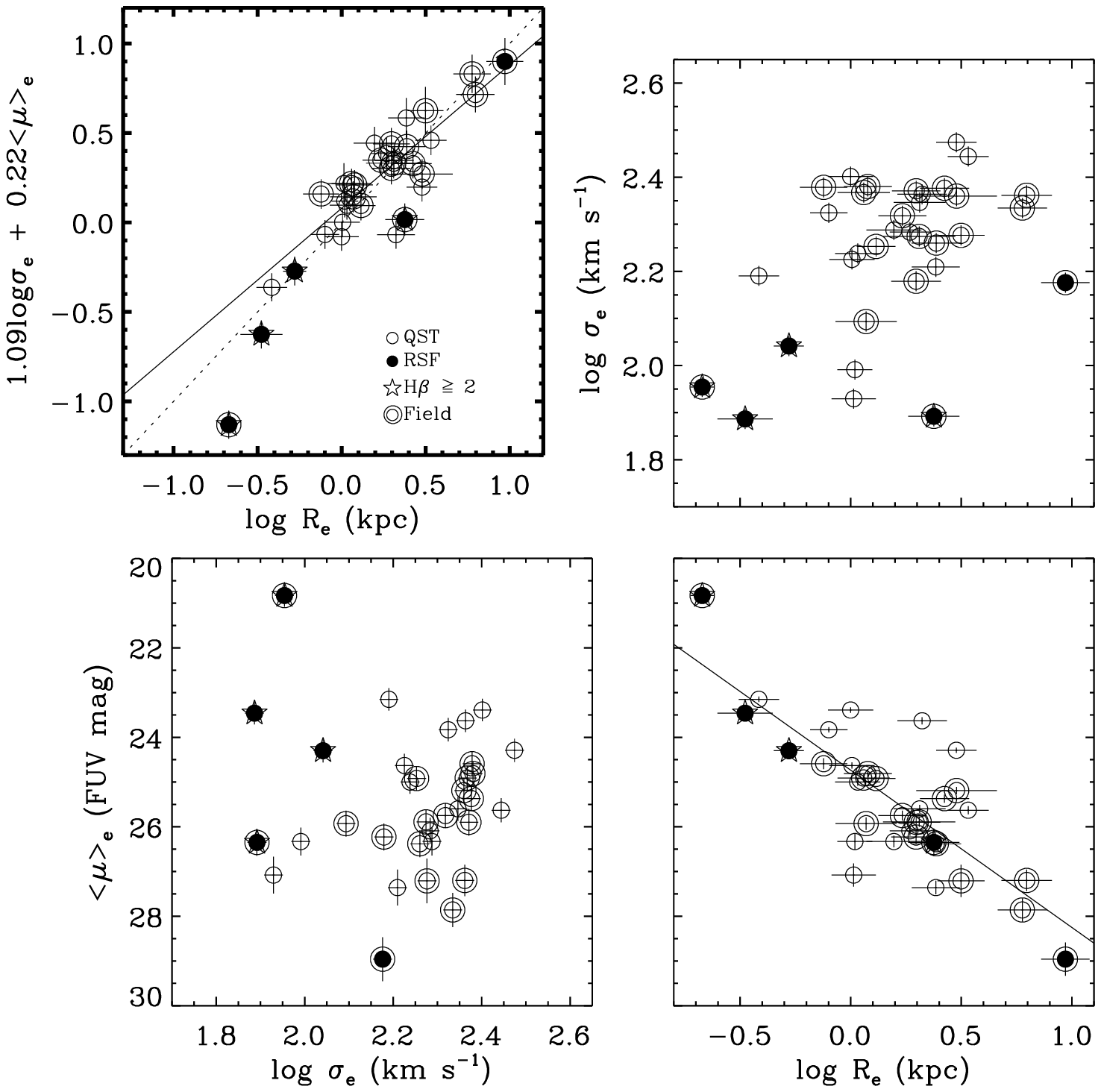}
\end{center}
\caption[]{Fundamental Plane and projections of the Fundamental
Plane in the FUV band. Symbols are the same as in
Figure~\ref{fig:CMRs} and linear fits to the two sub-samples (whole
sample, quiescent galaxies only) are shown as dotted and solid
lines, respectively. We also present a linear fit to the whole
sample as a solid line in the bottom-right plot (Kormendy
relation).}
\label{fig:fuvFP}
\end{figure*}
%

%
%
\begin{figure*}
\begin{center}
\includegraphics[width=14cm]{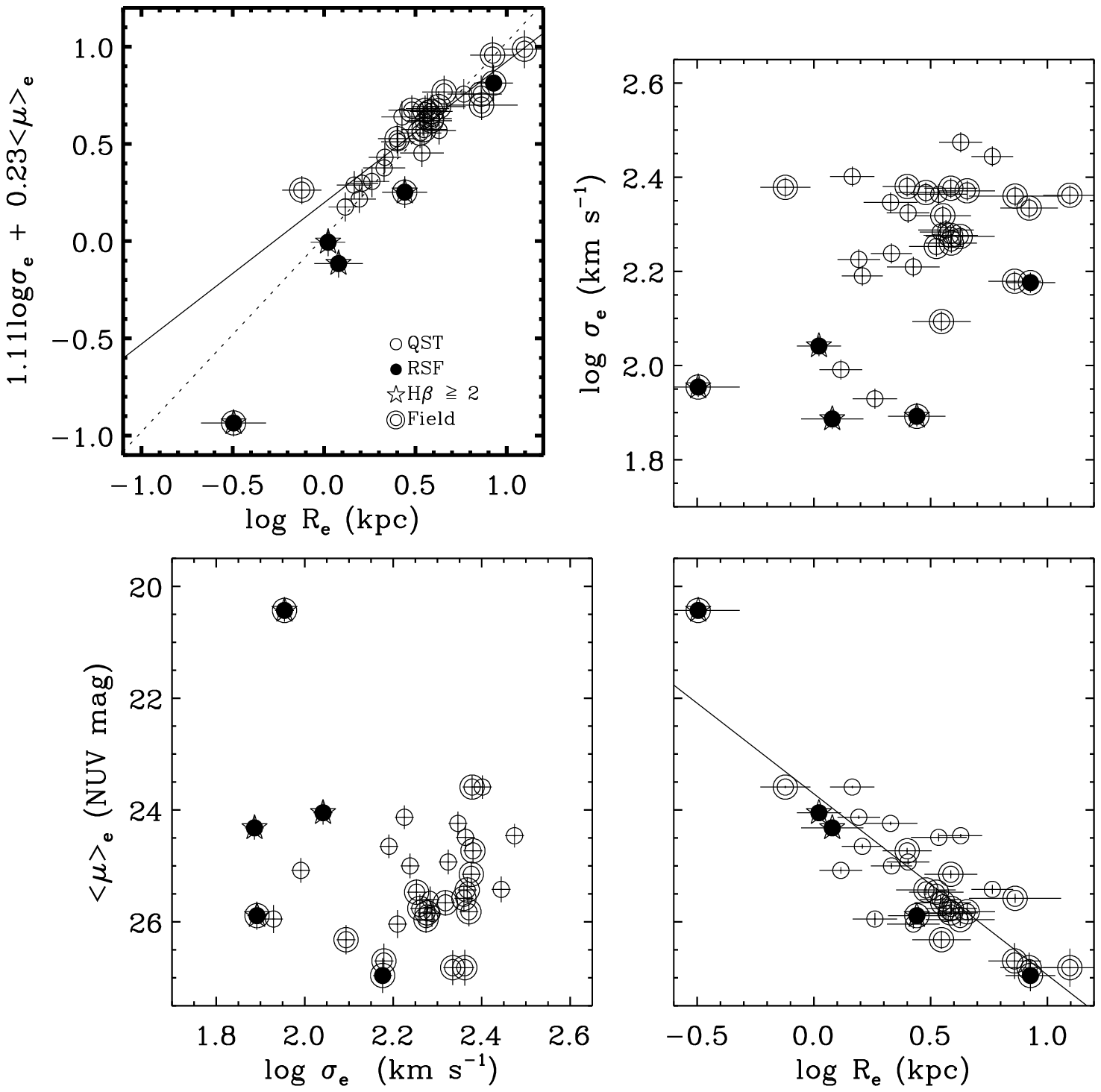}
\end{center}
\caption[]{Same as Figure~\ref{fig:fuvFP} but for the NUV band.}
\label{fig:nuvFP}
\end{figure*}
%

%
%
\begin{table}
\begin{center}
\caption{Fundamental Plane coefficients at FUV and NUV.}
\label{tab:FPresults}
\begin{tabular}{@{}lllll}
\hline
Band & $a$ & $b$ & Scatter & Population\\
\hline
FUV & 1.09 & 0.22 & 0.17 & All \\
FUV & 1.25 & 0.29 & 0.14 & QST \\
NUV & 1.11 & 0.23 & 0.14 & All \\
NUV & 1.82 & 0.30 & 0.08 & QST \\
\hline
\end{tabular}
\end{center}
\end{table}

For the first time, we present in Figures~\ref{fig:fuvFP} and
\ref{fig:nuvFP} the Fundamental Planes in the FUV and NUV bands,
using again the same symbols as in Figure~\ref{fig:CMRs}.
Table~\ref{tab:FPresults} lists the FP coefficients and scatter at
both FUV and NUV, obtained from a least-square fit by minimizing the
variations to the planes. Fits including and excluding RSF galaxies
are listed. As mentioned above, RSF galaxies tend to have lower
velocity dispersions and smaller effective radii and thus smaller
masses than the bulk of the sample galaxies. Crucially, RSF galaxies
systematically deviate from the best-fit planes so as to create
shallower slopes, and they significantly increase the scatters. This
is clearly shown by both Figures~\ref{fig:fuvFP} and \ref{fig:nuvFP}
and the FP parameters listed in Table~\ref{tab:FPresults}. Although
our sample is limited, this indicates that lower mass galaxies,
forming a larger amount of young stars, more greatly (and
systematically) deviate from the plane defined by the quiescent
galaxies.

The long-wavelength $K$-band slope is less sensitive to contamination
from young stellar populations than optical bands ($a\approx1.53$; see
\citealt{petal98a}). Considering our NUV Fundamental Plane fit for the
entire sample, our slope is significantly different from both the
virial expectation and the $K$-band value. However, comparing our
NUV-derived slope of $1.82$ for quiescent galaxies only, we recover
all of the $K$-band slope and almost all of the tilt away from the
virial expectation. We thus conclude that a significant fraction of
the FP tilt and scatter is due to low-mass early-type galaxies with
stellar populations significantly younger that those of high-mass
galaxies, thus reducing their mass-to-light ratios. The FUV
Fundamental Plane is more complicated to interpret because of the
additional presence of old but hot stars (UV upturn phenomenon), but a
similar if weaker trend is observed.

The tilt of the Fundamental Plane was also considered in
\citeauthor{cetal06} using $V$-band photometry and sophisticated
dynamical modeling. The main result to emerge was a clear,
continuous and monotonic increase of the total dynamical
mass-to-light ratio with velocity dispersion, consistent with the FP
tilt. As can be seen here in Figure~\ref{fig:CMRs2}, the NUV blue
galaxies affecting the tilt of the UV Fundamental Planes also have
some of the smallest velocity dispersions and lowest virial masses.
Furthermore, although there is non-negligible scatter, clear
(NUV$-V$)-$\sigma_{\rm e}$ and (FUV$-$NUV)-$\sigma_{\rm e}$ (or
(NUV$-V$)-$M_{\rm vir}$ and (FUV$-$NUV)-$M_{\rm vir}$) relations are
observed. The current results thus seem entirely consistent with
that of \citeauthor{cetal06}. We are simply picking up the galaxies
with the largest and/or most recent star formation events here with
our integrated NUV$-V\,<\,5.0$ criterion.

Many studies have also exploited the projection of the Fundamental
Plane onto the plane defined by $R_{\rm e}$ and
$\langle\mu\rangle_{\rm e}$, i.e.\ the Kormendy relation
\citep{k77}. A clear advantage of this relation is that it is based
on photometric parameters only, which are easily accessible even at
high redshifts. All of our sample galaxies follow the Kormendy
relation, and we note that our UV Kormendy relation slopes
($3.48\pm0.49$ and $3.24\pm0.36$ at FUV and NUV, respectively, see
Figures~\ref{fig:fuvFP} and \ref{fig:nuvFP}) are in rough agreement
with the optical one ($\langle\mu\rangle_{\rm
  e}\approx\,3.02\,\log\,R_{\rm e}$ in \citealt{hk87}). It may however
be significant that, compared to the mean trends, RSF galaxies tend to
be slightly offset toward smaller and/or fainter objects.
%
%
\subsection{UV radial colour profiles}
\label{sec:colour_profiles}
We use the radial colour profiles to discuss the stellar populations
of the sample galaxies. We discuss our criteria to classify them in
groups exhibiting similar properties below, and then discuss in more
detail galaxies with signs of recent star formation.

First, our main tool are the NUV$-V$ radial colour profiles. As for
integrated magnitudes, we classify a galaxy as having had recent star
formation if its NUV$-V$ profile is below $5.0$ for all or a
significant portion of the radii probed. Galaxies included in this
category are NGC~474, NGC~1023, NGC~2974, NGC~3032, NGC~4150,
NGC~4459, NGC~4526, NGC~4550 and NGC~7457. Having said that,
NUV$-$optical colours normally show negative slopes (the central
regions being redder) due to variations in the properties of the
underlying stellar populations, mainly age and metallicity. It is
well-known that there is a significant age-metallicity degeneracy
\citep[see, e.g.,][]{wo94}, whereby it is difficult to distinguish the
effects of a small change in age from those of a small change in
metallicity. Therefore, and even though we recognise the qualitative
aspect of such a criterion, if only a small part -- usually outside
the effective radius -- of the UV$-$optical colour profiles goes
smoothly and only a little below the NUV$-V\,=\,5.0$ cut, we consider
it as having been caused by a simple metallicity gradient rather than
residual star formation. In other words, localised regions with
NUV$-V\,<\,5.0$ are only classified as recently star-forming if the
changes in colour are abrupt. The galaxies exhibiting this behavior
include NGC~821, NGC~2699, NGC~4546, NGC~4570, NGC~5308, NGC~5813,
NGC~5831 and NGC~5845. Their integrated colours are all redder than
the NUV$-V\,=\,5.0$ colour threshold.

Second, we consider the FUV$-$NUV profiles. If a galaxy without
evidence of recent star formation has a central region with
FUV$-$NUV$\,<\,1.0$, we identify it as a UV-upturn galaxy. These
galaxies include NGC~4374, NGC~4486, NGC~4552, NGC4621 and NGC~5846.
As expected, these galaxies are among the most luminous and most
massive in our sample. All except NGC~4621 are slow-rotators (see
\citeauthor{eetal07}). Significantly, some UV-upturn galaxies have
extreme FUV$-$NUV and FUV$-V$ colours at small radii, bluer than the
integrated aperture colours of classic UV-upturn galaxies dating from
the {\it International Ultraviolet Explorer} ({\it IUE}) era.  As
these galaxies provided important constraints on stellar evolutionary
models of the UV-upturn, it will be interesting to revisit the models
with the current data. It is also probably worthwhile to point out
that a number of galaxies exhibit evidence for both a UV-upturn
(FUV$-$NUV$\,<\,1.0$ in the centre) and a substantial age or
metallicity gradient (NUV$-V\,<\,5.0$ in the outer parts). These
galaxies are NGC~2695, NGC~4278, NGC~4564, NGC~5198 and NGC~5982. They
probably harbour milder forms of the UV-upturn phenomenon, where
neither the UV-upturn nor large-scale stellar population gradients
dominates.

Finally, we note that some galaxies exhibit no evidence for recent
star formation or a UV-upturn, and no substantial large-scale age
and/or metallicity gradient. These galaxies are NGC~524, NGC~2768,
NGC~4387, NGC~4458, NGC~4473, NGC~4477 and NGC~5838.

Of the nine galaxies with convincing evidence for recent star
formation, five (NGC~474, NGC~1023, NGC~2974, NGC~4459 and NGC~4526)
show blue UV$-$optical colours in some regions only (as opposed to
all radii). Interestingly, this appears connected to kinematic
(sub-)structures detected in {\tt SAURON} and other studies. The
case of NGC~474, known for its shells \citep[e.g.][]{Tur99} and
strong misalignment between the kinematic and photometric major axes
(see \citeauthor{eetal04}), shows blue UV$-$optical colours and
enhanced H$\beta$ line strength at large radii. NGC~1023, with a
prominent twist in the centre of the velocity map
(\citeauthor{eetal04}), reveals a UV blob on the eastern side of the
galaxy and blue UV$-$optical colours in the same region, probably
related to an interacting gas-rich dwarf galaxy (see
\citealt{moetal06} for H\,{\small I} observations). The presence of
a UV-prominent outer ring related to a bar in NGC~2974 was discussed
by \citet{jetal07}. Blue UV$-$optical colours in this galaxy are
observed in the central and outer regions. In the case of NGC~4459,
blue UV$-$optical colours are observed only in the centre.
\citeauthor{ketal06} shows that this galaxy has strong H$\beta$
absorption within a central dust ring associated with a decoupled
stellar and ionised gas disc (\citeauthor{eetal04} and
\citeauthor{setal06}, respectively). NGC~4526 is a similar object
appearing more edge-on. Both NGC~4459 and NGC~4526 also harbor
central molecular gas discs \citep{ybc08}.

On the other hand, of the nine galaxies with recent star formation,
NGC~3032, NGC~4150, NGC~4550 and NGC~7457 show {\em overall} blue
UV$-$optical colours. NGC~3032 shows the bluest UV$-$optical colour
in our sample and has a prominent peak in the H$\beta$ map
suggesting recent and ongoing star formation. \citet*{mcetal06}
identified two types of kinematically-decoupled components (KDCs)
among the {\tt SAURON} sample: compact KDCs, which are often young
and occur in fast-rotating galaxies, and kiloparsec-scale KDCs, with
homogeneously old populations in slow-rotating galaxies (see also
\citeauthor{eetal07}). NGC~3032, NGC~4150 and NGC~7457 all have
small-scale KDCs (\citeauthor{eetal04}) and are considered fast
rotators (\citeauthor{eetal07}). NGC~3032 and NGC~4150 further have
substantial molecular gas discs \citep{ybc08}. In the case of
NGC~4550, with two counter-rotating stellar discs \citep[see,
e.g.,][]{rgk92,rietal92}, the H$\beta$ line strength is elevated in
the central region along the major axis (\citeauthor{ketal06}). Its
molecular gas content and detailed star formation history are
discussed in \citet{croetal08}. NGC~7457 has a controversial CO
detection discussed previously (see \citealt{ws03} and
\citealt{fc07}).
%
%
\section{Summary}
\label{sec:conclusions}
We have presented {\it GALEX} FUV and NUV imaging along with
ground-based F555W imaging from the MDM Observatory for $34$
early-type galaxies from the {\tt SAURON} survey sample. Nine of them
show extended blue UV$-$optical colours,
hinting at recent star formation. Five of these
are also classified as RSF
early-types by our integrated colour-magnitude relation
criterion. Supporting the findings from the UV colour-magnitude
relation technique, four out of the five candidate RSF galaxies also
show enhanced H$\beta$ absorption line strengths, a widely-used
post-starburst signature. The only exception (NGC~474) is noted for
its outer shells. Roughly $15$~per cent of the early-type galaxies in
our sample are therefore classified as RSF early types. Considering
that the UV flux from a starburst is only detectable for roughly
$1$~Gyr, this implies that residual star formation has been common in
early-type galaxies even up to the present day. This star formation
interpretation is also consistent with the results of molecular gas
surveys, as the CO detection rate is roughly $60$~per cent in RSF
early-type galaxies, much higher than in quiescent systems.

The velocity dispersions of these galaxies with evidence of recent
star formation are among the lowest in the sample, and none has
$\sigma_{\rm e}>200$~km~s$^{-1}$, consistent with the independent work
of \citet{scetal07}. Recent star formation early types also tend to
have smaller effective radii and thus smaller (virial) masses. Despite
the limited number statistics, a key result from the present study is
that RSF early types change the slopes of scaling relations
(colour-magnitude relations and Fundamental Planes) and dominate the
scatters in them. Most notably, much of the FP tilt and scatter can
now be explained by the fact that the properties of a substantial
fraction of early-type galaxies are influenced by RSF, systematically
biasing their mass-to-light ratios, especially at low masses. The UV
Fundamental Planes become significantly closer to the virial
expectation (and tighter) when only quiescent early-type galaxies are
considered. The same must be true at optical wavelengths, although the
effect will be smaller. Our result appears consistent with the
M/L\,-\,$\sigma_{\rm e}$ relation derived in \citeauthor{cetal06},
although the current colour threshold only picks up the most extreme
objects.

The radial UV$-$optical colour profiles not only reveal recent star
formation galaxies, but also a number of galaxies with smooth
large-scale age and/or metallicity gradients. Similarly, a number of
galaxies exhibiting the UV-upturn phenomenon are identified. Some
have bluer FUV$-$NUV and FUV$-V$ colours at small radii than the
integrated aperture colours of classic UV-upturn galaxies. A number
of galaxies show both a central UV-upturn and large-scale gradients.
This diversity of behaviours will, in due time, need to be explained
and reproduced by stellar population models.

Early-type galaxies are no longer thought to be simple. Many of them
have kinematic anomalies and sub-structures that are non-trivial to
interpret. We have shown here that they are also composed of
multiple generations of stars widely separated in age. This is
consistently found by other short-wavelength as well as far-infrared
studies, mainly aided by space experiments. The presence of young
stars in seemingly old populations is no longer debatable, yet it is
still unclear what causes and regulates this residual star
formation. We believe that our database, spatially
resolving nearby early-type galaxies, will be key to answering some
of those questions.
%
%
\section*{Acknowledgments}
The authors thank the anonymous referee for useful comments which led
to improvements in the paper. This work was supported by the Korea
Research Foundation Grant funded by the Korean government (KRF-C00156)
to SKY. MB acknowledges support from NASA through GALEX Guest
Investigator program GALEXGI04-0000-0109. MB and SKY are grateful to
the Royal Society for an International Joint Project award (2007/R2)
supporting this work. RLD acknowledges support from the Royal Society
in the form of a Wolfson Merit Award. MB and RLD are also grateful for
postdoctoral support through STFC rolling grant PP/E001114/1. The
STFC Visitors grant to Oxford also supported joint visits. GvdV
acknowledges support provided by NASA through Hubble Fellowship grant
HST-HF-01202.01-A awarded by the Space Telescope Science Institute,
which is operated by the Association of Universities for Research in
Astronomy, Inc., for NASA, under contract NAS 5-26555. MC
acknowledges support from a STFC Advanced Fellowship
(PP/D005574/1). Based on observations made with the NASA Galaxy
Evolution Explorer. GALEX is operated for NASA by the California
Institute of Technology under NASA contract NAS5-98034. Photometric
data were also obtained using the 1.3m McGraw-Hill Telescope of the
MDM Observatory. Part of this work is based on data obtained from the
ESO/ST-ECF Science Archive Facility. This project made use of the
HyperLeda database (http://leda.univ-lyon1.fr) and the NASA/IPAC
Extragalactic Database (NED) which is operated by the Jet Propulsion
Laboratory, California Institute of Technology, under contract with
the National Aeronautics and Space Administration.
%
%

%
\appendix
\section{Description for individual galaxies}
\label{sec:list}
We briefly comment here on the {\it GALEX} and MDM surface brightness
profiles of the E/S0 galaxies presented in this paper. Individual
comments on the stellar kinematics, the emission-line maps and the
line strength structures are given in \citeauthor{eetal04},
\citeauthor{setal06} and \citeauthor{ketal06}, respectively.
\begin{description}
\item[\bf NGC~474:] This galaxy (Arp~227), noted for its shell
  structures \citep[e.g.][]{Tur99}, shows blue NUV$-V$ colours except
  in the central regions. The H$\beta$ absorption line strength rises
  toward larger radii (\citeauthor[see][]{ketal06}). Considering the
  UV$-V$ colours, it seems that young stellar populations are present
  in the outer parts.
\item[\bf NGC~524:] This galaxy shows red colours in both the FUV$-V$
  and the NUV$-V$ colour profile, making a good example of a quiescent
  elliptical galaxy.
\item[\bf NGC~821:] This edge-on galaxy with a rapidly rotating
  disc-like component (\citeauthor[see][]{eetal04}) shows red FUV$-$NUV
  and UV$-V$ colours except outside the effective radius.
\item[\bf NGC~1023:] This SB0 galaxy shows a tilt in the FUV and NUV
  surface brightness profiles around $25$~arcsec, a central
  concentration in all metal lines (\citeauthor[see][]{ketal06}) and
  significantly negative values of both $h_3$ and $h_4$
  (\citeauthor[see][]{eetal04}). We discovered a strong NUV blob,
  constituting strong evidence for recent/ongoing star formation, just
  outside the galaxy. This is likely the result of a tidal interaction
  between NGC~1023 and a neighboring gas-rich galaxy
  \citep[see][]{moetal06}.
\item[\bf NGC~2695:] This galaxy shows blue FUV$-$NUV colours in the
  central regions with a high amplitude in $h_3$. The UV$-V$ colours
  become bluer toward larger radii, albeit with large
  uncertainties. We thus consider this a red galaxy.
\item[\bf NGC~2699:] The mean velocity map of this galaxy shows a
  rapidly rotating component around $5$~arcsec
  (\citeauthor[see][]{eetal04}). Similar to NGC~2695, it shows
  moderately blue NUV$-V$ colours only in the outer parts.
\item[\bf NGC~2768:] This galaxy with a cylindrical velocity field
  shows red colours in the UV$-V$ colour profiles.
\item[\bf NGC~2974:] This galaxy with a UV outer ring shows a blue
  colour in the outer regions, suggesting young stellar populations
  associated with a large-scale bar \citep[see][]{jetal07}, although
  the H$\beta$ line strength map appears relatively flat
  (\citeauthor[see][]{ketal06}).
\item[\bf NGC~3032:] Blue UV$-V$ colours are observed everywhere,
  suggesting a strong ongoing starburst. This galaxy also shows strong
  H$\beta$ absorption with negative Mg$b$ and Fe5270$_{\rm s}$ line
  strength gradients (\citeauthor[see][]{ketal06}).
\item[\bf NGC~4150:] This galaxy with a counter-rotating core
  (\citeauthor[see][]{eetal04}) shows blue UV$-V$ colours and enhanced
  H$\beta$ absorption with a strong drop in the Mg$b$ line strength
  (\citeauthor[see][]{ketal06}).
\item[\bf NGC~4278:] This galaxy shows blue FUV$-$NUV colours within
  $20$~arcsec, suggesting a strong FUV flux. It also has the strongest
  line emission (\citeauthor[see][]{setal06}).
\item[\bf NGC~4374 (M84):] This giant elliptical galaxy shows a
  similar FUV$-$NUV colour trend as that in NGC~4278. It is known for
  its BL~Lac nucleus and shows strong ionised-gas emission
  (\citeauthor[see][]{setal06}).
\item[\bf NGC~4387:] This boxy galaxy shows red UV$-V$ colours.
\item[\bf NGC~4458:] The FUV image of this galaxy exhibits an
  indistinct shape except for the central regions. This faint
  disc-like object also has a small central kinematically-decoupled
  core (\citeauthor[see][]{eetal04}).
\item[\bf NGC~4459:] This galaxy with a central dust ring
  (\citeauthor[see][]{setal06}) shows blue UV$-V$ colours within
  $10$~arcsec, coincident with an intense H$\beta$ absorption
  line strength and molecular gas.
\item[\bf NGC~4473:] This galaxy with a high velocity dispersion
  along the major axis likely resulting from substantial
  counter-rotation (\citeauthor[see][]{eetal04}) shows extended UV
  emission in the outskirts.
\item[\bf NGC~4477:] This galaxy with a prominent misalignment
  between the stellar kinematic and photometric major axes
  (\citeauthor[see][]{eetal04} shows overall red UV$-V$ colours.
\item[\bf NGC~4486 (M87):] This galaxy with a prominent jet shows
  blue UV$-V$ colours in the central regions, almost certainly related
  to nuclear activity. The jet also induces large errors in the surface
  brightness profiles.
\item[\bf NGC~4526:] The UV$-V$ plots of this galaxy exhibit blue
  colours in the centre, coincident with strong H$\beta$ absorption
  (\citeauthor[see][]{ketal06}) and a stellar velocity dispersion drop
  due to a star-forming fast-rotating disc
  (\citeauthor[see][]{eetal04} and \citeauthor{setal06}).
\item[\bf NGC~4546:] This galaxy shows moderately blue NUV$-V$
  colours in the outer parts. These could be related to the likely
  presence of a bar and associated star formation.
\item[\bf NGC~4550:] This galaxy with two counter-rotating stellar
  discs \citep[see][]{rgk92,rietal92} shows blue UV$-V$ colours
  everywhere.
\item[\bf NGC~4552 (M89):] This famous UV-upturn galaxy shows very
  blue FUV$-$NUV colours in the centre.
\item[\bf NGC~4564:] This elongated galaxy with a disc-like component
  (\citeauthor[see][]{eetal04}) shows blue FUV$-$NUV colours within
  the inner $10$~arcsec.
\item[\bf NGC~4570:] This fast-rotating edge-on galaxy shows slightly
  blue UV$-V$ colours in the outer parts. The presence of a bar was
  claimed by \citet{be98}.
\item[\bf NGC~4621:] This object has a disc-like component similar to
  NGC~4564 (\citeauthor[see][]{eetal04}). A counter-rotating component
  is also detected inside $2$~arcsec \citep{wec02}. Blue FUV$-$NUV
  colours are observed in the centre.
\item[\bf NGC~5198:] This galaxy with a central
  kinematically-decoupled core shows blue FUV$-$NUV colours in the
  centre that may indicate the presence of a UV upturn.
\item[\bf NGC~5308:] This edge-on disc galaxy shows red UV$-V$
  colours.
\item[\bf NGC~5813:] This galaxy with a well-known kinematically
  decoupled core shows marginally blue NUV$-V$ colours in the outer
  parts.
\item[\bf NGC~5831:] Like NGC~5813, this galaxy with a
  kinematically-decoupled core shows blue NUV$-V$ colours in the outer
  regions. There is a small bump in both FUV and NUV surface
  brightness profiles around $35$~arcsec.
\item[\bf NGC~5838:] This fast-rotating object shows red NUV$-V$
  colours throughout.
\item[\bf NGC~5845:] This compact elliptical galaxy shows a small
  bump in the NUV profile at a radius of $\approx10$~arcsec, with a
  corresponding distinct peak in the velocity map related to the
  central disk.
\item[\bf NGC~5846:] Similar to NGC~4552 and NGC~4564, this bright
  giant elliptical galaxy shows blue FUV$-$NUV colours in the centre
  that are suspected to be a UV upturn, with marginally blue NUV$-V$
  colours.
\item[\bf NGC~5982:] This galaxy with a kinematically-decoupled core
  shows blue FUV$-$NUV colours in the centre.
\item[\bf NGC~7457:] This kinematically-decoupled-core galaxy shows
  blue NUV$-V$ colours throughout, suggesting recent star
  formation. The H$\beta$ absorption is also high everywhere
  (\citeauthor[see][]{ketal06}).
\end{description}

\label{lastpage}
\end{document}